\documentclass[showpacs,preprintnumbers,
superscriptaddress,amsmath,floatfix]{revtex4}

\usepackage{graphicx}
\usepackage{bm} 

\newcommand{\non}{\nonumber\\}

\newcommand{\Diracslash}[1]{#1\llap{/\kern1pt}}
\newcommand{\diracslash}[1]{#1\llap{/\kern0pt}}

\newcommand{\be}{\begin{equation}}
\newcommand{\ee}{\end{equation}}
\newcommand{\bea}{\begin{eqnarray}}
\newcommand{\eea}{\end{eqnarray}}
\newcommand{\ba}[1]{\begin{array}{#1}}
\newcommand{\ea}{\end{array}}

\newcommand{\uk}{\hat{\mathbf{k}}}
\newcommand{\vk}{\mathbf{k}}
\newcommand{\vq}{\mathbf{q}}

\newcommand{\vp}{\mathbf{p}}

\newcommand{\vg}{\bm{\gamma}}

\begin{document}

\title{The Complex Gap in Color Superconductivity} 

\author{Philipp T.\ Reuter}
\email{reuter@triumf.ca}
\affiliation{
Institut f\"ur Theoretische Physik,
Johann Wolfgang Goethe-Universit\"at,
D-60054 Frankfurt, Germany}
\affiliation{
Department of Physics, University of Washington, 
Seattle, Washington 98195-1560, USA
}
\affiliation{
TRIUMF, 4004 Wesbrook Mall, Vancouver, BC, Canada, V6T 2A3
}

\date{\today} 

\begin{abstract}
We solve the gap equation for color-superconducting quark matter in the 2SC phase,
including both the energy and the momentum dependence of the gap, $\phi=\phi(k_0,\vk)$. For
that purpose a complex Ansatz for $\phi$ is made. 
The calculations are performed within an effective theory for cold and dense 
quark matter. The solution of the complex 
gap equation is valid to subleading order in the strong coupling constant $g$
and in the limit of zero temperature.
We find that, for momenta sufficiently close to the Fermi surface and for small energies, 
the dominant contribution to the imaginary part of $\phi$ arises from Landau-damped magnetic gluons.
Further away from the Fermi surface and for larger
 energies the other gluon sectors have to be included into  Im$\,\phi$.
We confirm that Im$\,\phi$ contributes a correction of order $g$ to the prefactor 
of $\phi$ for on-shell quasiquarks sufficiently close to the Fermi surface, whereas further away
from the Fermi surface Im$\,\phi$ and  Re$\,\phi$ are of the same order.
Finally, we discuss the relevance of Im$\,\phi$ for the damping of quasiquark excitations. 

\end{abstract}


\maketitle

\section{Introduction}

Sufficiently cold and dense quark matter is a color superconductor 
\cite{bailinlove,RWreview,alfordreview,schaferreview,DHRreview,Renreview,shovyreview,Huang}.
In the limit of asymptotically large quark chemical potentials, $\mu \gg \Lambda_{\rm QCD}$,
quarks are weakly coupled \cite{ColPer}, and interact mainly via single-gluon 
exchange. In this regime of weak coupling, the color-superconducting gap $\phi$
can be computed within the fundamental theory of strong interactions, quantum chromodynamics
(QCD) \cite{son,schaferwilczek,rdpdhr,Brown1,shovkovy,Hsu}. It was first noticed in Refs.\ 
\cite{son,schaferwilczek,rdpdhr} that in order to describe color superconductivity 
correctly it is crucial to take into account the specific energy and momentum dependence 
of the gluon propagator in dense quark matter. It turned out that the long-ranged,
magnetic gluons generate a logarithmic enhancement in addition to the standard
BCS logarithm and by that increase the value of the gap at leading logarithmic order.
Furthermore, due to the energy and momentum dependence of the gluon propagator 
the gap also is a function of energy and momentum and therefore a complex quantity
\cite{reuter2,ren}. Complex gap functions are well-known from the investigation of
strong-coupling superconductors in condensed matter physics for more than 40 years
 \cite{eliashberg,schrieffer,vidberg,mahan}.

The authors of Ref.\ \cite{rdpdhr} estimated the magnitude of the imaginary part of the 
color-superconducting gap for massless quarks by considering the cut of the
magnetic gluon propagator in the complex energy plane. 
They found that Im$\,\phi =0$ on the Fermi
surface and Im$\,\phi\sim g\,{\rm Re}\,\phi$ exponentially close to the Fermi surface,
$|k-\mu| \sim \mu \exp(-c/g)$ where $g\ll 1$ is the QCD coupling constant in the limit
of weak coupling. One only has ${\rm Im}\,\phi \sim {\rm Re}\,\phi$ for
quarks farther away from the Fermi surface, $|k-\mu| \sim g\mu$. 
Therefore, considering quarks exponentially close to the Fermi surface,
the approximation $\phi \simeq {\rm Re}\,\phi$ is valid
for $\phi$ up to corrections of order $g$ to the prefactor of $\phi$, i.e., 
up to its subleading order. 
However, the contribution of ${\rm Im}\,\phi$ to $\phi$ through ${\rm Re}\,\phi$
has not been estimated yet. 

In this work we show that, in the 2SC phase and to subleading order, 
${\rm Im}\,\phi$ does not contribute to $\phi$ for quarks exponentially close 
to the Fermi surface. To do so, all momentum and energy dependences
of $\phi$ must be included in the gap equation. The appropriate starting  point for
that is an energy- and momentum-dependent
Ansatz for $\phi$, $\phi = \phi(k_0,\vk)$, where $k_0$ and $\vk$ are treated as
independent variables. 
As known from the solution for ${\rm Re}\,\phi$, the integrals over energy and
momentum in the gap equation yield large logarithms, $\ln(\mu/\phi)\sim 1/g$,
which cancel powers of $g$ from the quark-gluon vertices. Due to these logarithms 
one must actually compute or at least carefully estimate these
integrals in order to determine the importance
of the various terms contributing to ${\rm Re}\,\phi$ and ${\rm Im}\,\phi$.
Moreover, for a complete account,
not only the magnetic cut but also the electric cut, as well as the poles of the
gluon propagator, have to be considered in the solution for ${\rm Im}\,\phi$. 
In order to illustrate the latter point, for energies just above the gluon mass
$m_g \sim g\mu$ one has  ${\rm Im}\,\phi \sim g^2\mu \gg {\rm Re}\,\phi$ due to a large 
contribution from the emission of on-shell electric gluons.
In order to estimate how this term feeds back into ${\rm Re}\,\phi$ requires a careful analysis.
Treating energy and momentum as independent variables and solving the coupled
 gap equations for Im$\,\phi$ and Re$\,\phi$ self-consistently is therefore a non-trivial problem 
 and, moreover, leads to interesting insights.

To date, the gap function has never been calculated by treating energy and momentum 
independently. In Refs.\ \cite{rdpdhr,meanfield,specgluon} it was assumed that the off-shell
gap function is of the same order of magnitude as the gap on the quasiquark mass-shell,
$\phi(k_0,\vk) \approx \phi(\epsilon_k,\vk) \equiv \phi_\vk$. 
Furthermore, all contributions that are generated by the energy dependence of the
gap, i.e., by its non-analyticities along the axis of real energies, are
neglected completely against the non-analyticities of the gluon propagator and 
the quark poles. In Ref.\ \cite{reuter} it was pointed out
that at least in order to calculate corrections of order $g$ to the prefactor of
the gap, its off-shell behaviour must be included. In Refs.\  \cite{son,RWreview,ren}, 
on the other hand, the gap has been calculated within the Eliashberg theory
\cite{mahan}. In this model it is assumed that Cooper pairing happens only on the
Fermi surface. Following this assumption, the external quark momentum in the gap
equation is approximated by  $k\approx \mu$. By additionally assuming isotropy in 
momentum space, the gap becomes completely independent of momentum and a function
of energy only, $\phi(\omega,\vk) \approx \phi(\omega,\mu) \equiv \phi(\omega)$.
Originally, the Eliashberg theory was formulated in order to include retardation
effects associated with the phonon interaction between electrons in a metal.
Since the energy that can be transfered between two electrons by a phonon
is restricted by the Debye frequency $\omega_D$,
Cooper pairing is restricted to happen only at the Fermi surface
\cite{mahan}.
In quark matter, however, such an assumption is problematic if
one is interested in understanding quark matter at more realistic densities where the 
coupling between the quarks becomes stronger. In this case, quarks further
from the Fermi surface also participate in the pairing \cite{Itakura}. 
If color superconductivity is present in the cores of neutron stars, the coupling is certainly strong,
$g\sim 1$, and a non-trivial momentum dependence of the gap cannot be neglected.
The present analysis is, strictly speaking, valid only at weak coupling. 
However, treating energy and momentum as independent variables might 
still be helpful to catch some aspects of color superconductivity at stronger coupling.

Besides affecting the value of the energy gap $\phi$ at some order, Im$\,\phi$ also 
contributes to the damping of the quasiquark excitations in a color superconductor.
With respect to the anomalous propagation of quasiquarks, it is shown that the imaginary
part of the gap broadens the support around the quasiquark poles, see Eq.\ (\ref{specxi}) below. 
This broadening strengthens the damping due to the imaginary part of the regular 
quark self-energy, $\Sigma$, which is present also in the non-colorsuperconducting medium
\cite{lebellac,vanderheyden,manuel2,manuel,rockefeller}.

This paper is organized as follows: In Sec.\ \ref{compgapeq} the 2SC gap equation
is set up within the effective theory derived in Ref.\ \cite{reuter}. In Sec.\ \ref{solving}
the gap equation is first decomposed into its real and imaginary parts and then solved 
to subleading order, cf.\ the schematic outline given after Eq.\ (\ref{traces}).
It is shown that Im$\,\phi$ 
contributes to $\phi$ at sub-subleading order for quarks exponentially close to the Fermi surface and
that  Im$\,\phi \sim$ Re$\,\phi$ at $|k-\mu| \sim g\mu$, which justifies previous calculations.
Furthermore, an analytical expression for Im$\,\phi$ is given, see Eq.\ (\ref{Imphiexact}) below.
In  Sec.\ \ref{outlook} the conclusions and an outlook are given.
A somewhat more detailed presentation can be found in Ref.\ \cite{reuter2}.

The units are $\hbar=c=k_B=1$. 4-vectors are denoted by
capital letters, $K^\mu = (k_0, {\bf k})$, with ${\bf k}$ being a
3-vector of modulus $|{\bf k}| \equiv k$ and direction
$\hat{\bf k}\equiv {\bf k}/k$. For the summation over Lorentz
indices,  we employ a metric
$g^{\mu \nu} = {\rm diag}(+,-,-,-)$ and perform the calculations
within a compact Euclidean space-time with volume $V/T$, where $V$
is the 3-volume and $T$ the temperature of the system.
Since space-time is compact, energy-momentum space is
discretized, with sums $(T/V)\sum_{K} \equiv T\sum_n (1/V) \sum_{\bf k}$. 
For a large 3-volume $V$, the sum over 3-momenta
can be approximated by an integral, $(1/V)\sum_{\bf k} \simeq
\int d^3 {\bf k}/(2 \pi)^3$. For bosons, the sum over $n$ runs over
the bosonic Matsubara frequencies $\omega_n^{\rm b} = 2n \pi T$, while 
for fermions, it runs over the fermionic Matsubara frequencies
$\omega_n^{\rm f} = (2 n+1)\pi T$.

\section{setting up the complex gap equation}\label{compgapeq}

The complex gap equation is set up within the effective theory derived in Ref.\ \cite{reuter}.
This has three major advantages over a treatment in full QCD: Firstly, self-consistency
of the solutions of the Dyson-Schwinger equations for the quark and gluon propagators is
only required for those momentum modes considered as {\em relevant\/} for the  physics of
interest. In full QCD, on the other hand, self-consistency has to be maintained for
{\em all\/} degrees of freedom. Secondly, by a special choice of the cutoffs for relevant
quarks and gluons, $\Lambda_{\rm q}$ and $\Lambda_{\rm gl}$, one can implement the
kinematics of quarks scattering along the Fermi surface into the effective theory. Considering
quarks with momenta $|k-\mu|<\Lambda_{\rm q}$ as relevant degrees of freedom, one can define
the projector onto these modes in Nambu-Gor'kov space as \cite{reuter}
\bea
{\cal P}_1(K,Q) & \equiv & \left( \begin{array}{cc}
 \Lambda_{\bf k}^+ & 0 \\
 0 & \Lambda_{\bf k}^- \end{array} \right) \, 
\Theta(\Lambda_{\rm q} - | k - k_F|) \, \delta^{(4)}_{K,Q}\;,
\eea
where $ \Lambda_{\bf k}^e\equiv(1+e\gamma_0 \vg\cdot \uk)/2$ projects onto states
 with positive ($e = +$) or negative ($e=-$) 
energy (quark masses being neglected). The quark modes far away from the Fermi surface as
well as antiquarks have the projector ${\cal P}_2 \equiv 1-{\cal P}_1$. They are integrated 
out and are contained in the couplings of the effective theory. For the gluons we introduce
the projector
\bea
{\cal Q}_1(P_1,P_2) & \equiv & \Theta(\Lambda_{\rm gl} -p_1) \, \delta^{(4)}_{P_1,P_2}\;.
\eea
Consequently, relevant gluons are those with 3-momenta less than $\Lambda_{\rm gl}$, while
gluons with larger momenta, corresponding to ${\cal Q}_2 \equiv 1-{\cal Q}_1$,
are integrated out. Choosing the cutoffs according to
\bea\label{cutoffs}
\Lambda_{\rm q}  \alt g \mu \ll \Lambda_{\rm gl} \alt \mu
\eea
the energy of a gluon exchanged between two quarks is restricted by $p_0 < \Lambda_{\rm q}$.
Its momentum, on the other hand, can be much larger, since $p < \Lambda_{\rm gl}$.
This reflects the fact that quarks typically scatter along the Fermi surface and, due to the Pauli
principle, do not penetrate deeply into the Fermi sea. In addition to that, gluons with $p_0 \ll p$
have the property that they are not screened in the magnetic sector and therefore dominate 
the interaction among quarks. The third advantage of this effective theory is that  by expanding
 the numerous terms in the gap equation in terms of 
$\Lambda_{\rm q} /\Lambda_{\rm gl} \sim g $ one can systematically identify contributions of
leading, subleading, and sub-subleading order. This was demonstrated explicitely in 
\cite{reuter} for the real part of the gap equation. Similarly, also the terms in the complex  
gap equation can be organized in this way.
%
Obviously, the separation of the scales $\phi\,,g\mu$, and $\mu$ is rigorously valid only 
at asymptotically large values of the quark chemical potential, where $g \ll 1$. In the 
physically relevant region, $\mu \alt 1$ GeV and $g \sim 1$, this scale hierarchy breaks down.
For that case, more suitable choices for cutoff parameters
have been suggested \cite{schaferschwenzer}.

The Dyson-Schwinger equation for relevant quarks and gluons can be derived in a systematic way
using the Cornwall-Jackiw-Tomboulis (CJT) formalism \cite{CJT}.
For the quarks one finds
\be \label{NGinvquark}
{\cal G}^{-1} =   \left( \begin{array}{cc}
                            [ G^+]^{-1} &  0 \\
                     0 & [ G^-]^{-1} \end{array} \right)
                + \left( \begin{array}{cc}
                            \Sigma^+ &  \Phi^- \\
                     \Phi^+ & \Sigma^- \end{array} \right) \;.
\ee
Here $[G^+]^{-1}$ is the inverse tree-level propagator for quarks and $[G^-]^{-1}$ is the
corresponding one for charge-conjugate quarks. These effective propagators differ from the QCD 
tree-level propagator $[G_0^\pm]^{-1}(K) \equiv \Diracslash{K} \pm \mu \gamma_0$ by additional
loops of irrelevant quark and gluon propagators. In Ref.\ \cite{reuter} it is shown that to subleading
order in the gap equation these loops can be neglected, $[G^\pm]^{-1}\simeq [G_0^\pm]^{-1}$.
The regular self-energy for (charge-conjugate) quarks is denoted as $\Sigma^\pm$. The off-diagonal 
self-energies $\Phi^\pm$, the gap matrices, connect regular with charge-conjugate
quark degrees of freedom. A non-zero $\Phi^\pm$ corresponds to the condensation of quark Cooper
pairs. Equation (\ref{NGinvquark}) can be formally solved for ${\cal G}$,
\be \label{NGquarkprop}
{\cal G} \equiv \left( \begin{array}{cc}
                            {\cal G}^+ &  \Xi^- \\
               \Xi^+ & {\cal G}^- \end{array} \right) \; ,
\ee
where
\be
{\cal G}^\pm \equiv \left\{ [G^\pm]^{-1} + \Sigma^\pm - 
\Phi^\mp \left( [G^\mp]^{-1} + \Sigma^\mp \right)^{-1} \Phi^\pm \right\}^{-1}
\ee
is the propagator describing normal propagation of quasiparticles
and their charge-conjugate counterpart, while 
\be \label{Xi}
\Xi^\pm \equiv - \left( [G^\mp]^{-1} + \Sigma^\mp \right)^{-1}
\Phi^\pm {\cal G}^\pm
\ee
describes anomalous propagation of quasiparticles, which is possible if the ground state is
a color-superconducting quark-quark condensate, for details, see Ref.\ \cite{DHRreview}.
To subleading order, it is sufficient to approximate the propagator of the soft gluons by the 
HDL-resummed propagator $\Delta_{\rm HDL}$ instead of solving the corresponding
 Dyson-Schwinger equation \cite{dirkselfenergy}, while for the hard gluons one may 
 use the free propagator $\Delta_{0,22}$ \cite{reuter}. The index 22 indicates that this
  propagator describes the propagation of a hard gluon mode. One has in total
\bea\label{splitgluonprop}
\Delta^{\mu \nu}_{ab}(P) \equiv 
 \left[ \Delta_{\rm HDL}\right]^{\mu \nu}_{ab}(P)\,\theta(\Lambda_{\rm gl}-p) +
\left[ \Delta_{0,22}\right]^{\mu \nu}_{ab}(P) \,\theta(p-\Lambda_{\rm gl})\;.
\eea
In the mean-field approximation \cite{meanfield} the
Dyson-Schwinger equation for the gap matrix $\Phi^+ (K)$ reads
\be\label{gapequation1}
\Phi^+ (K)  =  g^2 \, \frac{T}{V} \sum_Q 
\Delta^{\mu \nu}_{ab}(K-Q)
 \, \gamma_\mu (T^a)^T \, \Xi^+(Q) \, \gamma_\nu T^b   \;,
\ee
cf.\ Eq.\ (97) in Ref.\ \cite{reuter}. As discussed above, in the effective theory the sum runs only 
over relevant quark momenta, $\mu - \Lambda_{\rm q} \leq q \leq \mu + \Lambda_{\rm q}$. Due to
the dependence of the gluon propagators $\Delta_{\rm HDL}$ and $\Delta_{0,22}$ 
on the external quark energy momentum $K$ in Eq.\ (\ref{gapequation1}), 
the solution $\Phi(K)^+$ 
must be energy-dependent itself. Hence, solving the gap equation self-consistently requires an 
energy-dependent Ansatz for the gap function. To subleading order in the gap equation, the 
contribution from the regular self-energies $\Sigma^{\pm}$ can be subsumed by replacing
$q_0 \rightarrow q_0/Z(k_0)$ in the quark propagators \cite{qwdhr}, where
\be \label{wavefunc}
Z(k_0) = \left( 1 + \bar{g}^2  \,\ln \frac{M^2}{k_0^2} \right)^{-1}
\ee 
is the quark wave-function renormalization factor \cite{manuel,rockefeller}, with
\bea\label{gbar}
\bar g\equiv \frac{g}{3\sqrt{2}\pi}\;,
\eea
and
\bea \label{Mconst}
M^2\equiv \frac{3\pi}{4} \,m_g^2\;,~~~m_g^2 \equiv N_f \frac{g^2\mu^2}{6\pi^2}\;.
\eea
The effect of Im$\,\Sigma$ on Re$\,\phi$ has been studied in Ref.\ \cite{manuel2}, where it is
shown that Im$\,\Sigma$ suppresses the formation of quark Cooper pairs. The corresponding 
corrections, however, are shown to enter Re$\,\phi$ only beyond subleading order. In the 
following it will be assumed that Im$\,\Sigma$ enters Re$\,\phi$ through Im$\,\phi$ also only 
beyond subleading order. Consequently, Im$\,\Sigma$ will be neglected completely. 
This is self-consistent since it turns out that Im$\,\phi$ itself contributes
only beyond subleading order to Re$\,\phi$.
The main contributions to Im$\,\phi$ are expected to arise from the energy dependence of
the gluon propagator, and not from Im$\,\Sigma$. This amounts
in neglecting the cut of the logarithm in Eq.\ (\ref{wavefunc}) when performing the Matsubara sum
in the complex gap equation (\ref{gapequation1}).

For the sake of definiteness, a two-flavor color superconductor is considered, where the
 color-flavor-spin structure of the gap matrix is \cite{DHRreview,reuter}
\be \label{gapmatrix}
\Phi^+(K) = J_3 \tau_2 \gamma_5\, \Lambda_{\bf k}^+ \, 
\Theta(\Lambda_q -|k-\mu|)\, \phi(K)\;.
\ee
The matrices $(J_3)_{ij} \equiv -i \epsilon_{ij3}$ and $(\tau_2)_{fg} \equiv-i \epsilon_{fg}$
represent the fact that quark pairs condense in the color-antitriplet, flavor-singlet channel.
Then the anomalous propagator reads
\be\label{Xi2}
\Xi^+(Q) = J_3 \tau_2 \gamma_5 \, \Lambda_{\bf q}^- \,
\Theta(\Lambda_q - |q-\mu|) \, \frac{\phi(Q)}{[q_0/Z(q_0)]^2 - \epsilon_q^2}\;,
\ee
where
\bea\label{gapexcite}
{\epsilon_\vk} = \sqrt{(k-\mu)^2 + \phi^2}\;.
\eea
Here we employed the analytical continuation  $|\phi|^2 \rightarrow \phi^2$ 
\cite{ren,schrieffer,mahan}. Besides its poles at
$q_0 =\pm Z(\epsilon_q)\epsilon_q \equiv \pm \tilde \epsilon_q$ the anomalous propagator
$\Xi^+$ obtains further non-analyticities along the real $q_0$-axis through the complex gap 
function $\phi$.
As presented more explicitly in Appendix \ref{disper}, the energy dependence of the gap function 
$\phi(K)$ gives rise to a non-trivial spectral density
\bea\label{specphi1}
\rho_\phi(\omega,\vk) &\equiv &  \frac{1}{2\pi i}\,\left[\phi(\omega+i\epsilon,\vk)-
				\phi(\omega-i\epsilon,\vk)\right]\;,
\eea
which is directly related to the imaginary part of the gap function via
\bea\label{imphidef}
{\rm Im}\,\phi(\omega+i\epsilon,\vk)&=& {\pi}\,\rho_\phi(\omega,\vk)\;.
\eea
For the spectral density of the anomalous quark propagator this yields
\bea\label{specxi}
\rho_\Xi(\omega,\vq)&\equiv& 
\frac{1}{2\pi i}\left[  \Xi(\omega+i\eta,\vq) - \Xi(\omega-i\eta,\vq) \right]\non
&=&
- Z^2(\omega){\mathcal P}\; \frac{\rho_{\phi}(\omega,\vq)}{\omega^2-[Z(\omega)\epsilon_\vq]^2} - 
{\rm sign}(\omega)\,Z^2(\tilde\epsilon_\vq)\,{\rm Re}\,\phi(\omega+i\eta,\vq)\,\delta\!
\left(\omega^2 -\tilde\epsilon_\vq^2 \right)\;,
\eea
where the cut of $Z(\omega)$ has been neglected. Also the non-analyticities of $\epsilon_q$
can be neglected: In the region $|k-\mu|\sim \phi$ one has $\phi\approx $ Re$\,\phi$, 
whereas for $|q-\mu|\sim g\,\mu \gg \phi$ it is ${\epsilon_q}\simeq |q-\mu|$. Hence, in Eq.\ (\ref{specxi})
and in the gap equation (\ref{gapequation1}) one may write 
${\epsilon_q} \simeq \sqrt{(q-\mu)^2 + ({\rm Re}\,\phi)^2}$, which is continuous across the 
real energy axis. From Eq.\ (\ref{specxi}) it becomes obvious that $\rho_{\phi}(\omega,\vq)\neq 0$
leads to a broadening of $\rho_\Xi(\omega,\vq)$ around the quasiparticle pole at
 $\omega \equiv \tilde\epsilon_q$. However, in order to 
describe the damping of quasiquarks self-consistently, it is necessary to include Im$\,\Sigma$.
Further interesting details of the damping due to Im$\,\phi$ can be found in the recent analysis
 of Ref.\  \cite{ren}

Inserting Eq.\ (\ref{Xi2}) into Eq.\ (\ref{gapequation1}), multiplying from both sides with
$J_3 \tau_2 \gamma_5 \Lambda_{\bf k}^+$,
and tracing over color, flavor, and Dirac degrees of freedom, one finds with
$[\Delta_{0,22}]^{\mu \nu}_{ab} \equiv \delta_{ab}\, \Delta_{0,22}^{\mu
\nu}$ and $[\Delta_{\rm HDL}]^{\mu \nu}_{ab} \equiv \delta_{ab}\, 
\Delta_{\rm HDL}^{\mu \nu}$
 \bea \label{gapequation2a}
\phi(K) = \frac{g^2}{3} \, \frac{T}{V} \sum_Q 
{\rm Tr}_s \left( \Lambda_{\bf k}^+ \gamma_\mu 
\Lambda_{\bf q}^- \gamma_\nu \right)
 \Delta^{\mu \nu}(K-Q)\,\tilde\Delta(Q)\,\phi(Q)\;,
\eea
where the remaining traces run over the internal Dirac indices. For convenience, the quark
 propagator
\bea\label{quarkpropcompact}
\tilde\Delta(Q)\equiv \frac{Z^2(q_0)}{q_0^2 - [Z(q_0)\,\epsilon_q]^2} =
\frac{1}{2\tilde\epsilon_\vq}\sum\limits_{\sigma=\pm}
\frac{\sigma\,Z^2(q_0)}{q_0 -\sigma Z(q_0)\,\epsilon_\vq}
\eea 
has been introduced. Before the actual solution of Eq.\ (\ref{gapequation2a}) in the 
next section, we first discuss the power-counting scheme of the gap equation in weak coupling. 
In order to fulfill the equality in Eq.\ (\ref{gapequation2a}), the integration 
over energy and momentum on the r.h.s.\ must yield terms of the order $\phi/g^2$.
 After combination with the prefactor $g^2$ they are of order $\phi$, which is 
the leading order in the gap equation. Accordingly, terms of order $g\,\phi$ are of 
subleading order and terms of order $g^2\phi$ are of sub-subleading order. Until now, the 
color-superconducting gap is known up to subleading order, i.e., up to corrections of order $g$ 
to the prefactor of the gap \cite{DHRreview}.

\section{Solving the complex gap equation}\label{solving}

\subsection{Derivation of the coupled gap equations of Re$\,\phi$ and Im$\,\phi$}\label{anacont}

\begin{figure}[ht]
\centerline{\includegraphics[width=6cm]{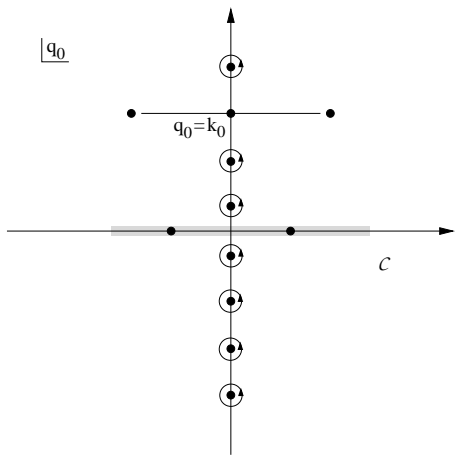}}
\caption{The contour ${\cal C}$ in Eq.\ (\ref{Mdef})
encloses the poles of $\tanh [q_0/(2T)]$ on the
imaginary $q_0$ axis. The additional poles and the cut at $q_0=k_0$ arise from the gluon 
propagator, while the two poles on the real axis are due to the quasiquarks, cf. Eq.\ (\ref{specxi}).
The yet undetermined non-analyticities of the gap function on the real $q_0-$axis are indicated
 by the shaded area.}
\label{phicontour1}
\end{figure}

\begin{figure}[ht]
\centerline{\includegraphics[width=6cm]{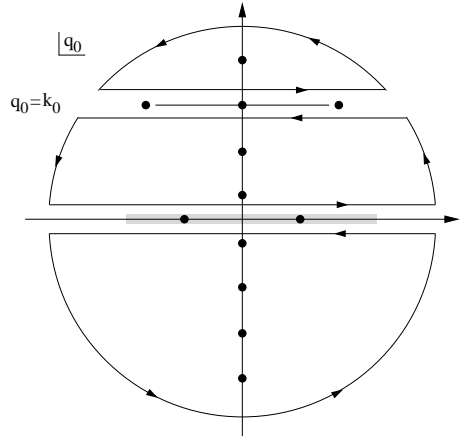}}
\caption[Deforming the contour ${\cal C}$.]{Deforming the contour ${\cal C}$ to introduce the spectral
densities of $\Delta^{\ell,t}$ and $\phi$, cf.\ Eq.\ (\ref{M1}).}
\label{phicontour2}
\end{figure}

In order to derive of the coupled gap equations of Re$\,\phi$ and Im$\,\phi$ we
first rewrite the Matsubara sum in Eq.\ (\ref{gapequation2a}) as a contour integral, 
\bea\label{Mdef}
{\mathcal M}^{\ell,t}(k_0,\vp,\vq) \equiv
 T \sum\limits_{q_0\neq k_0}\Delta^{\ell,t}(Q-K)\tilde\Delta(Q)\phi(Q)
=\int\limits_{\mathcal C} \frac{dq_0}{2\pi i}\,
\frac{1}{2}\tanh\left(\frac{q_0}{2T} \right)\Delta^{\ell,t}(Q-K)\tilde\Delta(Q)\phi(Q) \;,
\eea
where $\vp = \vq-\vk$ and $\Delta^{\ell}$ denotes the longitudinal and $\Delta^{t}$ the
transverse gluon propagator in pure Coulomb gauge, cf.\ App.\ \ref{specgluons}. 
The contribution at $q_0 = k_0$,
where the cut of the gluon propagator is located, has to be omitted.
The contour $\mathcal C$ is shown in cf.\ Fig.\ \ref{phicontour1}.
In order to introduce the spectral densities of the gap function and of the magnetic 
and longitudinal gluon propagators the contour ${\cal C}$ 
is deformed corresponding to Fig.\ \ref{phicontour2}.
The contour integral can now be 
decomposed into three parts
\bea\label{M}
{\mathcal M}^{\ell,t}(k_0,\vp,\vq)= I_\infty^{\ell,t} + I_0^{\ell,t} + I_{k_0}^{\ell,t}\;.
\label{M1}
\eea
The first part, $I_\infty$, is the integral along a circle of asymptotically large radius. It can 
be estimated after parameterizing $dq_0 = i|q_0|e^{i\theta}d\theta$ and considering the limit 
$|q_0|\rightarrow \infty$. To this end we write without loss of generality, cf.\ Appendix \ref{disper},
\bea\label{phisplit}
\phi(K)\equiv \tilde\phi(K) +{\hat\phi}(\vk)\;,
\eea
so that the energy dependence of $\phi(K)$ is contained in $ \tilde\phi(K)$. For asymptotically large
energies, $|k_0|\rightarrow \infty$, we require that $\tilde\phi(K) \rightarrow 0 $ and 
$\phi(K) \rightarrow {\hat\phi}(\vk)$, where ${\hat\phi}(\vk)$ is a real function of $\vk$ only.
 Furthermore, we have $\Delta^\ell(Q-K) \rightarrow |\vq-\vk|^{-2}$ 
and $\Delta^t(Q-K) \rightarrow -q_0^{-2}$, cf.\ Eq.\ (\ref{longtransprops4}). It follows that the 
longitudinal contribution $I_\infty^{\ell}\sim q_0^{-1}$ and the transversal 
$I_\infty^{t} \sim q_0^{-3}$, 
and hence that in total $I_\infty^{\ell,t} \rightarrow 0$.
 The integral $I_0^{\ell,t}$ runs around the axis of real $q_0$,
\bea \label{I00}
I_0^{\ell,t}=
\int\limits_{-\infty}^\infty \frac{dq_0}{2\pi i}\,
\frac{1}{2}\tanh\left(\frac{q_0}{2T} \right)\Delta^{\ell,t}(Q-K)
\left[\tilde\Delta(q_0+i\eta,\vq)\phi(q_0+i\eta,\vq)-
\tilde\Delta(q_0-i\eta,\vq)\phi(q_0-i\eta,\vq)\right]\;.
\eea
Applying the Dirac identity and using  Eq.\ (\ref{specphi1}) 
and $\phi(\omega +i \eta)+\phi(\omega-i\eta) = 2{\rm Re}\,\phi(\omega)$ one obtains
\bea\label{I0}
I_0^{\ell,t}
&=&
\frac{1}{2{\tilde \epsilon_\vq}}\sum\limits_{\sigma=\pm}\sigma{\mathcal P}_{\sigma{\tilde \epsilon_\vq}} 
\int\limits_{-\infty}^\infty dq_0 \frac{1}{2}\tanh\left(\frac{q_0}{2T}\right)
\Delta^{\ell,t}(q_0-k_0,\vp)Z^2(q_0)\frac{\rho_{\phi}(q_0,\vq)}{q_0-\sigma{\tilde \epsilon_\vq}}\non
&&-\frac{1}{2{\tilde \epsilon_\vq}}\frac{1}{2}\tanh\left(\frac{{\tilde \epsilon_\vq}}{2T}\right)
Z^2(\tilde\epsilon_\vq)\, {\rm Re}\,\phi({\tilde \epsilon_\vq},\vq) \sum\limits_{\sigma=\pm}
\Delta^{\ell,t}(\sigma{\tilde \epsilon_\vq}-k_0,\vp)\;,
\eea
where ${\mathcal P}_x$ denotes the principal value with respect to the pole at $x$.
The first term arises from the non-analyticities of the gap function, cf.\ Eq.\ (\ref{specphi1}),
and the second from the poles of the quark propagator at $q_0 = \pm {\tilde \epsilon_\vq}$, 
cf.\ Eq.\ (\ref{quarkpropcompact}). 

The last integral $I_{k_0}^{\ell,t}$ circumvents the non-analyticities
 of the gluon propagator as well as the pole of $\tanh$. One finds
\bea\label{Ik0}
I_{k_0}&=&\frac{1}{2{\tilde \epsilon_\vq}}\sum\limits_{\sigma=\pm}\sigma{\mathcal P}_{0} 
\int\limits_{-\infty}^\infty dq_0 \frac{1}{2}\coth\left(\frac{q_0}{2T}\right)\phi(q_0+k_0,\vq)\,
Z^2(q_0+k_0)\,\frac{\rho^{\ell,t}(q_0,\vp)}{q_0-\sigma{\tilde \epsilon_\vq}+k_0}\non
&& -T\, \Delta^{\ell,t}(0,\vp) \,\tilde\Delta(k_0,\vq)\,\phi(k_0,\vq)\;.
\eea
The first term is due to the non-analyticities of the longitudinal and magnetic gluon propagators,
 $\Delta^{\ell,t}$. The second arises from the pole of $\coth(q_0/2T)$ at $q_0=0$ and corresponds
to the large occupation number density of gluons in the classical limit, $q_0 \ll T$. This 
contribution has been shown to be beyond subleading order \cite{rdpdhr} and will therefore be
discarded in the following.

After analytical continuation, $k_0 \rightarrow \omega + i\eta$, the Dirac identity is employed
in order to split the complex gap equation (\ref{gapequation2a}) into its real and imaginary part.
For its imaginary part one finds using Eqs.\ (\ref{I0},\ref{Ik0}) and using the fact  that 
Re$\,\phi(q_0)$ and $Z^2(q_0)$ are even functions
\bea\label{ImM}
{\rm Im \mathcal M}^{\ell,t}(\omega+i\eta,\vp,\vq) &=&
-\frac{\pi}{4{\tilde \epsilon_\vq}}{\rm Re}\,\phi({\tilde \epsilon_\vq},\vq)\,
Z^2(\tilde\epsilon_\vq)\sum\limits_{\sigma=\pm}\sigma\rho^{\ell,t}
({\omega-\sigma\tilde \epsilon_\vq},\vp)
\left[\tanh\left(\frac{\sigma{\tilde \epsilon_\vq}}{2T}\right)
 +\coth\left(\frac{{\omega-\sigma\tilde \epsilon_\vq}}{2T}\right) \right]\nonumber\\
&&+
\frac{\pi}{4{\tilde \epsilon_\vq}}\sum\limits_{\sigma=\pm}\sigma{\mathcal P} 
\int\limits_{-\infty}^\infty dq_0
\frac{\rho^{\ell,t}(\omega-q_0,\vp)\rho_{\phi}(q_0,\vq)}{q_0-\sigma{\tilde \epsilon_\vq}}
Z^2(q_0)\left[\tanh\left(\frac{q_0}{2T}\right) +\coth\left(\frac{\omega-q_0}{2T}\right) \right] 
\non
&\equiv& {\rm Im \mathcal M}^{\ell,t}_{\cal A}(\omega+i\eta,\vp,\vq)+
{\rm Im \mathcal M}^{\ell,t}_{\cal B}(\omega+i\eta,\vp,\vq)\,.
\eea
The first term on the r.h.s\ of Eq.\ (\ref{ImM}), ${\rm Im \mathcal M}^{\ell,t}_{\cal A}$, 
contains all contributions to $\phi(K)$ that have already been considered in previous solutions 
 to subleading order. Here the gap function is always on the quasiparticle mass-shell.
The second term, ${\rm Im \mathcal M}^{\ell,t}_{\cal B}$, is due to the non-analyticities
of $\phi(K)$. It has been neglected in all previous treatments
due to the approximation
\bea\label{ansatz}
\rho_\Xi(\omega,\vq)\simeq  -{\rm sign}(\omega)\,{\rm Re}\,\phi(\tilde\epsilon_\vq+i\eta,\vq)
\,Z^2(\tilde\epsilon_\vq)\,\delta\!\left(\omega^2 -\tilde\epsilon_\vq^2 \right)\;,
\eea
cf. Eq.\ (41) in Ref.\ \cite{rdpdhr}. By doing so, one always forces the gap function 
on the r.h.s.\ of the gap equation on the quasiparticle mass-shell.
The gap equation takes then the standard form, cf.\ Eq.\ (19) of Ref.\ \cite{qwdhr}.
The occurrence of the external energy $\tilde\epsilon_k$ on the r.h.s.\ due to the 
energy-dependent gluon propagators indicates that the solution still possesses 
some energy dependence although not provided by the Ansatz.

In the limit of small temperatures, $T\rightarrow 0$, the hyperbolic functions 
in Eq.\ (\ref{ImM}) simplify, yielding for $\omega > 0$
\bea\label{ImMT0}
{\rm Im \mathcal M}^{\ell,t}_{T=0}(\omega+i\eta,\vp,\vq) &=&
\frac{\pi}{2{\tilde \epsilon_\vq}}\left[\frac{}{}
\!\!-Z^2(\tilde\epsilon_\vq)\,{\rm Re}\,\phi({\tilde \epsilon_\vq},\vq)\,\rho^{\ell,t}
(\omega-{\tilde \epsilon_\vq},\vp)\,\theta(\omega-{\tilde \epsilon_\vq})\right.\non
&&\left.
+\sum\limits_{\sigma=\pm}\sigma\,{\mathcal P} \int\limits_{0}^{\omega} dq_0
\;\frac{\rho^{\ell,t}(\omega-q_0,\vp)\rho_{\phi}(q_0,\vq)}{q_0-\sigma{\tilde \epsilon_\vq}}
\,Z^2(q_0)\right]\\
&\equiv& {\rm Im \mathcal M}^{\ell,t}_{{\cal A},T=0}(\omega+i\eta,\vp,\vq)+
{\rm Im \mathcal M}^{\ell,t}_{{\cal B},T=0}(\omega+i\eta,\vp,\vq)\;.
\eea
Inserting the Matsubara sums ${\rm Im\mathcal M}^{\ell,t}_{\cal A,B}$ back into 
Eq.\ (\ref{gapequation2a}),
\bea\label{Imphieq}
{\rm Im}\,\phi(\omega+i\eta, \vk)&=&
\frac{g^2}{3}\int\frac{d^3q}{(2\pi)^3} 
\sum \limits _{r=\ell,t}{\rm Tr}_s^r ( k,p, q)\,\left[
{\rm Im\mathcal M}^r_{\cal A}(\omega+i\eta,\vp,\vq)+
{\rm Im\mathcal M}^r_{\cal B}(\omega+i\eta,\vp,\vq)
\right]\\
&\equiv& {\cal A}(\omega+i\eta, \vk)+ {\cal B}(\omega+i\eta, \vk)\label{AB}
\;,
\eea
we have finally derived the gap equation for Im$\,\phi(\omega+i\eta, \vk)$.
The traces over Dirac space are given by
\begin{subequations} \label{traces}
\bea
{\rm Tr}_s^{\ell} ( k,p, q)& = & \frac{(k+q)^2 - p^2}{2\, k\,q} \;
, \\
{\rm Tr}_s^{t} ( k,p, q) & = & -2 - \frac{p^2}{2\, k\,q} + 
\frac{(k^2-q^2)^2}{2\, k\, q\, p^2}\;.
\eea
\end{subequations}

We found that the imaginary part of the gap function can be split into a term
 ${\cal A}(\omega+i\eta, \vk)$ which  contains all contributions 
of the real part of the gap function on the quasiquark mass-shell,
and a term ${\cal B}(\omega+i\eta, \vk)$ which contains all new contributions
with the imaginary part of the gap function off the quasiparticle mass-shell, 
cf.\ Eq.\ (\ref{AB}).
In the following it will be checked whether the known solution for 
Re$\,\phi(\tilde\epsilon_\vk,\vk)$, which neglects the contributions contained in 
$\cal B$, is self-consistent to subleading order. To this end, we can use the 
leading order solution for \mbox{Re$\,\phi(\tilde\epsilon_\vk,\vk)$}, which is given by
\cite{DHRreview}
\begin{equation} \label{solution}
\phi(y) \equiv \phi \, \sin\left( \frac{\pi\,y}{2}\right)\,\, ,
\end{equation}
where  $\phi$ denotes the value of the gap function on the Fermi surface to leading 
logarithmic  order in $g$ \cite{son}
\be \label{phileading}
\phi \sim \mu \exp\left(-\frac{\pi}{2\, \bar{g}}\right) \;.
\ee
The variable $0\leq y \leq 1$ defines the distance  from the Fermi surface through the mixed scale
\bea\label{Lambday}
\Lambda_y \equiv \phi^y M^{1-y}\;,
\eea
where $M\sim g\mu$ is defined in Eq.\ (\ref{Mconst}). A given value of  $y$ corresponds to
 momenta $\vk$ with $|k-\mu|\sim \Lambda_{y}$.  Note that $\Lambda_1=\phi$ and 
 $\Lambda_0=M$. Furthermore, we have $\Lambda_{\bar g} \sim e^{-\pi/2}\,M$, which is smaller
 but still of the order of $M$.
 Correspondingly, we refer to quarks with momenta $|k-\mu| \sim \Lambda_{1\leq y<\bar g}$ as 
 {\it exponentially close} to the Fermi surface and to quarks with $|k-\mu| \sim \Lambda_{\bar g\geq y\geq 0}$ 
 as {\it farther away} from the Fermi surface.
 Inserting $\phi(y)$ into ${\cal A}(\omega+i\eta, \vk)$ one can estimate 
${\cal A}(\omega+i\eta, \vk)$ for different energy and momentum regimes. This is done in 
Sec.\ \ref{estA}. In 
the second iteration the part ${\cal B}(\omega+i\eta, \vk)$
is estimated by inserting $\rho_\phi\simeq {\cal A}/\pi$ into the 
expression for ${\cal B}(\omega+i\eta, \vk)$, which is done in Sec. \ref{estB}.
In Sec.\ \ref{hilbert} these estimates are used to write
\bea\label{splitting}
{\rm Re}\,\tilde\phi(\epsilon_\vk,\vk) =
\frac{1}{\pi}\mathcal P\left[\;\int\limits_0^{\Lambda_1}+
\int\limits _{\Lambda_1}^{\Lambda_{\bar g}}+
\int\limits _{\Lambda_{\bar g}}^{\Lambda_0}+\cdots\right] d\omega \,\sum\limits_{\sigma=\pm}
\frac{{\cal A}(\omega+i\eta,\vk)+{\cal B}(\omega+i\eta,\vk)}{\omega-\sigma\epsilon_\vk}\;,
\eea
cf.\ Eq.\ (\ref{disp1}), 
where the integral over $\omega$ has been split according to the different energy 
regimes of the estimates for $\cal A$ and $\cal B$, cf.\ Table \ref{tablesummary}.
Then, according to the discussion after Eq.\ (\ref{quarkpropcompact}),
terms of order $g^n\phi$ contribute to the (sub)$^n$-leading order
to ${\rm Re}\,\tilde\phi(\epsilon_\vk,\vk)$. The main results of this analysis are summarized 
in Table \ref{tablesummary} for momenta close to the Fermi surface, $|k-\mu|\ll M$.
\begin{table}  
\centerline{\begin{tabular}[t]{|c||c|c|c|c|c|}
\hline
  & ~$\omega\lesssim\Lambda_1$~ &~ $\omega \sim\Lambda_{1>y>\bar g}$~ &
  ~ $\omega\sim \Lambda_{\bar g>y>0}$ ~&
~ $\omega\sim m_g +\Lambda_{1>y>0}$~ &~ $M\lesssim\omega\lesssim 2\mu$ ~
\\ \hline\hline
~dominant gluons~  & $t$-cut	& $t$-cut & $t,\ell$-cut & $\ell$-pole &  $t$-pole \\ \hline
~ ${\cal A}$~  &  ~ $g^2\phi$ ~ & ~ $g\,\phi\,\cos\left(\frac{\pi\,y}{2}\right)$  ~  & 
 $g\,\phi$  &~ $g\,\phi\left(\frac{M}{\phi}\right)^y$  ~  &  $g\,\phi$   \\ \hline
~ ${\cal B}$~  &  ~ $g^2{\cal A}$ ~ & ~$g^2{\cal A}$~  &  $g{\cal A}$  &
~  $g^2{\cal A}$  ~  &  $g{\cal A}$ \\ \hline
~ ${\cal H}\,[{\cal A}]$~  &  ~ $g^2\phi$ ~ & ~$\phi$~  &  $g\,\phi$  &
~  $\phi$ ~  &  $g\,\phi$  \\ \hline
~ ${\cal H}\,[{\cal B}]$~  &  ~ $g^4\phi$ ~ & ~$g^2\phi$~  &  $g^2\phi$  &
~  $g^2\phi$ ~  &  $g^2\phi$  \\ \hline
\end{tabular}}
\caption{Estimates for the terms ${\cal A}$ and  ${\cal B}$, cf.\ Eq.\ (\ref{AB}),
and for ${\cal H}\,[{\cal A}]$ and ${\cal H}\,[{\cal B}]$,  cf.\ Eq.\ (\ref{splitting}),
for different energy scales and $|k-\mu|\ll M$. The gluon sectors dominating the 
respective energies are indicated.}
\label{tablesummary}
\end{table}
The columns correspond to the various energy regimes of these estimates. In the first
 line the dominant gluon sectors are given. 
The cut of the transversal gluons gives the dominant contribution to ${\cal A}$ and $ {\cal B}$  
for energies smaller than the scale $M$. At the scale $M$, the longitudinal and the transversal cut
contribute with the same magnitude. At that energy scale also the poles of the gluons start to 
contribute as soon as $\omega > m_g$. At energies just above $m_g$ the longitudinal pole dominates 
over  the  magnetic pole, 
whereas for larger energies up to $2\mu$ the transversal pole gives the leading contribution.
The respective orders of magnitude of ${\cal A}$ and $ {\cal B}$, estimated at the various
 energy scales, are given in the subsequent rows. 
 It is found that either ${\cal B} \sim g {\cal A}$ or ${\cal B} \sim g^2 {\cal A}$ and therefore ${\cal B \ll \cal A}$. 
 Finally, the orders of magnitudes of the contributions  to the Hilbert transforms
${\cal H}\,[{\cal A}]$ and ${\cal H}\,[{\cal B}]$, obtained by integrating over the respective energy scales,
 cf.\ Eq.\ (\ref{splitting}), are listed. When calculating the Hilbert transform 
${\cal H}\,[{\cal A}]$ up to energies $\omega \sim \Lambda_{\bar g}$,
the transversal cut gives a contribution of order $\phi$, which is of leading order.
Since ${\cal B}\sim g^2{\cal A}$ for these energies, ${\cal H}\,[{\cal B}] \sim g^2\phi$, i.e., 
the contributions from $\cal B$ are beyond subleading order.  
Integrating over larger energies $\omega \sim M$, it turns out that ${\cal H}\,[{\cal A}] \sim g\,\phi$, 
which gives a subleading-order contribution. Since ${\cal B}\sim g{\cal A}$ in this energy regime, the
corresponding contribution from ${\cal H}\,[{\cal B}]$ is again beyond subleading order.
For the contributions from the poles one finds that, in the region 
$\omega = m_g  +\Lambda_{1>y>0}$, a contribution of order $\phi$ is generated by ${\cal H}\,[{\cal A}]$
(which combines with $\hat\phi(\vk)$ to give a contribution of order $g\,\phi$ in total, i.e., of subleading order).
Since in this energy regime ${\cal B}\sim g^2{\cal A}$, the corresponding contribution 
 ${\cal H}\,[{\cal B}]$ is again only of order $g^2 \phi$, i.e., beyond subleading order. 
Integrating over large energies up to $2\mu$, ${\cal H}\,[{\cal A}]$ gives a contribution of order $g\,\phi$, which 
is of subleading order. Since in this regime ${\cal B}\sim g{\cal A}$, it follows that ${\cal H}\,[{\cal B}]$ is
beyond subleading order.

In Sec.\ \ref{phi0} it is shown that also ${\hat\phi}(\vk)\sim \phi$ and that the imaginary part of the 
gap function again contributes to  ${\hat\phi}(\vk)$ only at $g^2\phi$, i.e., beyond subleading order. 
This finally proves that the imaginary part of the gap function enters the real part only beyond subleading order. 
Hence, to subleading accuracy, the real part of the gap equation can be solved self-consistently neglecting the 
imaginary part of the gap function for quark momenta exponentially close to the Fermi surface. 
On the other hand, the real part of the gap function enters the imaginary part of the gap function always
to leading order. For energies, for which  ${\cal B}\sim g^2{\cal A}$,  the imaginary part of the 
gap function can be calculated  to subleading order from the real part alone. In particular, this is the case for
the regime of small energies. Furthermore, since for these energies only the cut of magnetic gluons contributes,
Im$\,\phi$ can be calculated to subleading order without much effort, which is done in Sec.\ \ref{calcimphi}.
In Sec.\ \ref{repro}, we reproduce the known gap equation for Re$\,\phi(\tilde\epsilon_\vk,\vk)$ by 
Hilbert transforming  ${\cal A}(\omega+i\eta, \vk)$, cf.\ Eq.\ (\ref{disp1}),
and adding the energy independent gap function $\hat\phi(\vk)$, cf.\ Eq.\ (\ref{phisplit}).

\subsection{Estimating ${\cal A}$ }\label{estA}  

For the purpose of estimating the various terms contributing to $\cal A$, cf.\  Eqs.\ 
(\ref{ImMT0}-\ref{AB}), one may restrict oneself to the leading
 contribution of the Dirac traces in Eq.\ (\ref{gapequation2a}), which is of order one. The integral over the 
 absolute magnitude of the quark momentum is $\int dq \, q^2$, while the angular integration is 
$\int d \cos \theta \equiv \int d p \, p / (kq)$. Furthermore, we estimate $Z^2(\tilde\epsilon_\vq)\sim 1$.
The contribution to $\cal A$
from ${\cal M}^{\ell,t}_{{\cal A},T=0}$ in Eq.\ (\ref{ImMT0}),
which arises from the cut of the soft gluon propagator, is
\bea\label{Acut}
{\cal A}^{\ell,t}_{\rm cut}(\omega,\vk)&\sim&g^2\int\limits_0^\delta\frac{d\xi}{\epsilon_\vq}\,{\rm Re}\,
\phi(\epsilon_\vq,\vq)\int\limits_\lambda^{\Lambda_{\rm gl}} dp\, p \,\rho^{\ell,t}_{\rm cut}(\omega^*,\vp)\;,
\eea
where $\omega^*\equiv\omega-\epsilon_\vq$ and $0<\omega^*<\omega$. Furthermore, 
 $\delta\equiv{{\rm min}(\omega,\Lambda_{\rm q})}$ and
  $\lambda\equiv {\max}(|\xi -\zeta|,\omega^*)$
with $\zeta \equiv k-\mu$ and $\xi \equiv q-\mu$. In the effective theory,
 both $|\zeta|$ and $|\xi|$ are bounded by
$\Lambda_{\rm q} \sim g\mu$. Due to the condition $\lambda<p<\Lambda_{\rm gl}$
 it follows that ${\cal A}^{\ell,t}_{\rm cut}= 0$ for 
 $\omega >\Lambda_{\rm gl}+\Lambda_{\rm q}\simeq \mu$. For 
the purpose of power counting it  is sufficient to employ the following
 approximative forms for
$\rho^{\ell,t}_{\rm cut}$, 
cf.\ Eq.\ (\ref{lcut},\ref{tcut}),
\begin{subequations} \label{appcut}
\bea \label{appcutt}
\rho_{\rm cut}^t (\omega^*, {\bf p}) &\simeq& \frac{M^2}{\pi} \,
\frac{\omega^* \, p}{p^6 + (M^2\, \omega^*)^2}\;,\\
\rho^{\ell}_{\rm cut}(\omega^*,{\bf p}) &\simeq&
\frac{2 M^2}{\pi}\, \frac{\omega^*}{p}\,\frac{1}{
( p^2 + 3\, m_g^2 )^2} \;. \label{appcutl}
\eea
\end{subequations}
These approximations reproduce the correct behavior for $\omega^*\ll p\ll m_g$, while for
$\omega^* < p < \Lambda_{\rm gl}$ they give at least the right order of magnitude.  
With that the integration over $p$ can be performed analytically. 
For energies $\omega <\Lambda_{\rm gl}$ one finds for the transverse part
\bea\label{t}
{\cal A}^t_{\rm cut}(\omega,\vk)&\sim&g^2\int\limits_0^{\delta}\frac{d\xi}{\epsilon_\vq}
\,{\rm Re}\,\phi(\epsilon_\vq,\vq)
\left[\arctan\left(\frac{\Lambda_{\rm gl}^3}{M^2\omega^*}\right)-\arctan\left(\frac{\lambda^3}
{M^2\omega^*}\right)\right]
\;.
\eea
For all $\zeta \leq \Lambda_{\rm q}$ and $\omega \leq \Lambda_{\rm gl}$ one has
$\Lambda_{\rm gl}^3/M^2\omega^*\gg 1$ and the first arctangent in the squared brackets
may be set equal to $\pi/2$. In the case that $\omega \gg M$ one has
 $\lambda^3/M^2\omega^*\gg 1$ and the two arctangents cancel. If $\omega \sim M$, on the other
hand, the arctangents combine to a number of order $1$. Finally, in the case that 
$\omega \sim \Lambda_{y}$ with $0\leq y \leq 1$, one has always $0\leq \xi\leq \Lambda_{y}$ due to the
theta-function in Eq.\ (\ref{ImMT0}). Moreover, it turns out that the arctangents cancel if 
$ \zeta > \Lambda_{y/3}$, since then
 $\lambda^3/(M^2\omega^*)\simeq \zeta^3/[M^2 (\omega-\xi)] \gg1$ for all 
 $\xi <\omega$.  

In the case that $\omega\sim\phi$, the integral over $\xi$ does not yield the BCS logarithm,
 cf.\ Appendix \ref{nonBCS}, and we find
\bea\label{Aphi}
{\cal A}^t_{\rm cut}(\phi,\vk)&\sim&g^2\phi\;.
\eea
For larger energies $\omega\sim\Lambda_y$ with $0\leq y<1$ one substitutes
$\xi(y^\prime)\equiv \Lambda_{y^\prime}$, $d\xi/\xi = \ln(\phi/M)\,dy^\prime$ and obtains 
with Eq.\ (\ref{solution})
\bea
{\cal A}^t_{\rm cut}(\omega,\vk)&\sim& g^2\,\ln\left(\frac{\phi}{M}\right) \phi\int\limits_1^y dy\,
\sin\left( \frac{\pi\,y}{2}\right)\sim  g\,\phi\,\cos\left( \frac{\pi\,y}{2}\right)\;.\label{Aphi2}
\eea
Here, the integration over $0\leq \xi \leq \Lambda_1$ has been neglected, since it
 gives at most a contribution of order
$g^2\phi$, cf.\ Eq.\ (\ref{Aphi}).

In the longitudinal sector one finds for the integral over the gluon momentum $p$ 
\bea\label{pintlong}
{\cal I}(\lambda)&\equiv& M^2\int\limits_\lambda^{\Lambda_{\rm gl}} \frac{dp}{
( p^2 + X^2 )^2}\sim \frac{1}{X}\left[\arctan\left(\frac{\Lambda_{\rm gl}}{X}\right)-
\arctan\left(\frac{\lambda}{X}\right)\right]
 - \frac{1}{\Lambda_{\rm gl} }\frac{\lambda^2-\lambda\Lambda_{\rm gl}+X^2}{\lambda^2+X^2}\non
&\sim&
\left\{
\begin{array}{c}
\frac{1}{X}~~~~~~,~{\rm for}~\lambda \leq X\\
\frac{\Lambda_{\rm gl}-\lambda}{\Lambda_{\rm gl}\lambda}~~,~{\rm for}~\lambda \gg X
\end{array}
\right.
\;,
\eea
where $X^2\equiv 3m_g^2$. Since $\zeta \leq \Lambda_{\rm q} \sim X$, solely the magnitude of $\omega$ decides 
whether $\lambda\leq X$ or $\lambda\gg X$ is realized. It follows that, in contrast to the transversal case, 
the order of magnitude of ${\cal A}^\ell_{\rm cut}$ is independent of $\zeta$. Energies 
$\omega \sim \Lambda_{0\leq y\leq 1}$ correspond to $\lambda \leq X$. For the special case
$\omega \sim \Lambda_1$ one finds
\bea
{\cal A}^\ell_{\rm cut}(\omega,\vk)&\sim&g^2\int\limits_0^{\Lambda_1}\frac{d\xi}{\epsilon_\vq}\,
{\rm Re}\,\phi(\epsilon_\vq,\vq)\,\,\frac{\omega^*}{M} \sim g^2\,\phi\,\frac{\phi}{M}\;.
\eea
For $\omega \sim \Lambda_{0\leq y< 1}$ one obtains
\bea\label{Aphi2b}
{\cal A}^\ell_{\rm cut}(\omega,\vk)&\sim&g\,\phi\int\limits_1^y dy^\prime\,\sin\left( \frac{\pi\,y^\prime}{2}\right)\,
\frac{\omega^*}{M} 
\sim g\,\phi\int\limits_1^y dy^\prime\,\sin\left( \frac{\pi\,y^\prime}{2}\right)\,\left[\left(\frac{\phi}{M} \right)^y
-\left(\frac{\phi}{M} \right)^{y^\prime}\right]\non
&\sim& g\,\phi\,\cos\left( \frac{\pi\,y}{2}\right) \,\left(\frac{\phi}{M} \right)^y\;.
\eea
Hence, for $y>0$ the term ${\cal A}^\ell_{\rm cut}$ is suppressed by a factor $(\phi/M)^y$ as compared 
to ${\cal A}^t_{\rm cut}$ whereas for  $y=0$ the longitudinal and the transversal cut contribute at the same order,
  ${\cal A}^\ell_{\rm cut} \sim {\cal A}^t_{\rm cut}\sim g\,\phi$.
For much larger energies, $M \ll \omega < \mu$,  we have $\lambda=\omega^* \simeq \omega \gg X$
 (note that $\zeta$ is bounded by $\Lambda_{\rm q}$). It follows with
 $\delta = \Lambda_{\rm q} \sim \Lambda_0$
\bea
{\cal A}^\ell_{\rm cut}(\omega,\vk)&\sim&g\,\phi\int\limits_1^0 dy\,\sin\left( \frac{\pi\,y}{2}\right)\,
\left(1-\frac{\omega}{\Lambda_{\rm gl}}\right) 
\sim g\,\phi \left(1-\frac{\omega}{\mu}\right) \;. \label{Acuthigh}
\eea
Hence, ${\cal A}^\ell_{\rm cut} \gg {\cal A}^t_{\rm cut}$ in this large-energy regime. The results for 
${\cal A}^{\ell,t}_{\rm cut}$ are summarized in Tables \ref{tableAcutzeta<M} and \ref{tableAcutzetasimM}.

\begin{table}  
\centerline{\begin{tabular}[t]{|c||c|c|c|c|}
\hline
 & $\omega\sim\phi$ & $\omega\sim\Lambda_{1>y>0}$ & ~$\omega\sim M$~ & ~$M\ll\omega<\mu$~
\\ \hline\hline
~ ${\cal A}_{\rm cut}^t$ ~  &  $g^2\phi$	& $g\,\phi\,\cos\left(\frac{\pi\,y}{2}\right)$ & $g\,\phi$ & 0
\\ \hline
${\cal A}_{\rm cut}^\ell$  & ~ $g^2\phi\,\frac{\phi}{M}$~ 	&
 ~$g\,\phi\left(\frac{\phi}{M}\right)^y\cos\left(\frac{\pi\,y}{2}\right)$
 ~ &  $g\,\phi$  & ~$g\,\phi\left(1-\frac{\omega}{\mu}\right)$~
\\ \hline
\end{tabular}}
\caption{Estimates for ${\cal A}_{\rm cut}^{\ell,t}$ at different energy scales and $\zeta\ll M$.}
\label{tableAcutzeta<M}
\end{table}

\begin{table}  
\centerline{\begin{tabular}[t]{|c||c|c|c|c|}
\hline
  & $\omega\sim\phi$ & $\omega\sim\Lambda_{1>y>0}$ & ~$\omega\sim M$~ & ~$M\ll\omega<\mu$~
\\ \hline\hline
~ ${\cal A}_{\rm cut}^t$ ~  &  0 & 0  & $g\,\phi$ & 0
\\ \hline
${\cal A}_{\rm cut}^\ell$  & ~ $g^2\phi\,\frac{\phi}{M}$~ 	&
 ~$g\,\phi\left(\frac{\phi}{M}\right)^y\cos\left(\frac{\pi\,y}{2}\right)$
 ~ &  $g\,\phi$  & $~g\,\phi\left(1-\frac{\omega}{\mu}\right)$~
\\ \hline
\end{tabular}}
\caption{Estimates for ${\cal A}_{\rm cut}^{\ell,t}$ at different energy scales
 and $\zeta\lesssim M$.}
\label{tableAcutzetasimM}
\end{table}

\begin{figure}[ht]
\centerline{\includegraphics[width=6cm]{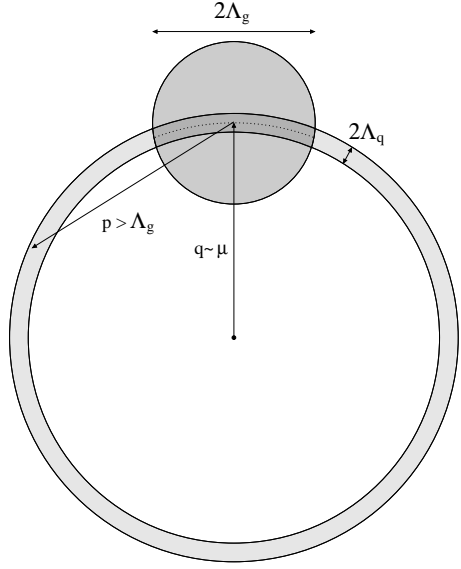}}
\caption{Hard gluon exchange with momentum $p>\Lambda_{\rm gl}\sim \mu$. The quark has 
to remain within a layer of width $2\Lambda_{\rm q}\sim g\mu$ around the Fermi surface.
 This effectively restricts the hard gluon momentum to 
 $p< 2\mu+2\Lambda_{\rm q}\lesssim 2\mu$.}
\label{Sphere3}
\end{figure}

The contributions from the gluon poles to $\cal A$ read analogously to Eq.\ (\ref{Acut})
\bea
{\cal A}^{\ell,t}_{\rm pole}(\omega,\vk)&\sim&
g^2\int\limits_0^\delta\frac{d\xi}{\epsilon_\vq}\,{\rm Re}
\,\phi(\epsilon_\vq,\vq)\int\limits_{|\xi-\zeta|}^{2\mu}
 dp\, p \,\rho^{\ell,t}_{\rm pole}(\omega^*,\vp)\,
\delta[\omega^*-\omega_{\ell,t}(\vp)]\;.\label{Apole}
\eea
The boundary $p \leq  2\mu$ in the integral over $p$ is due to the constraint
 $\xi\leq\Lambda_{\rm q}$, cf.\ Fig.\ \ref{Sphere3}.
Therefore, also $\omega_{\ell,t}<2\mu$. It follows that ${\cal A}^{\ell,t}_{\rm pole}=0$ for 
$\omega> 2\mu$, since for these energies one has $\omega^*\simeq \omega> \omega_{\ell,t}$ 
always and the $\delta-$function in Eq.\ (\ref{Apole}) vanishes. Hence, we can restrict the analysis
of Eq.\  (\ref{Apole}) to $m_g < \omega < 2\mu$. 

For the transverse sector we approximate
\bea\label{approxt}
\rho^{t}_{\rm pole}(\omega_{t}(\vp),\vp)\simeq-\frac{1}{2\omega_t(\vp)}
\eea
and $\omega_{t}(\vp)\simeq \sqrt{p^2+m_g^2}$ for all momenta $p$ and obtain
\bea
{\cal A}^{t}_{\rm pole}(\omega,\vk)&\sim&
g^2\int\limits_0^{\Lambda_{\rm q}}\frac{d\xi}{\epsilon_\vq}\,{\rm Re}\,\phi(\epsilon_\vq,\vq)\!\!\!\!
\int\limits_{\sqrt{m_g^2+|\zeta-\xi|^2}}^{2\mu}\!\!\!\!\!\!d\omega_t\,\delta(\omega^*-\omega_{t})\;.
\eea
The condition $\omega^*= \omega_t$ can be satisfied only if $\omega >\sqrt{m_g^2+\zeta^2}$. 
Furthermore, the condition $\omega^*= \omega_t$ sets an upper boundary
to the integral over $\xi$ given by
$\xi_{\rm max} \equiv {\rm min}\{\Lambda_{\rm q}, (\omega^2-m_g^2-\zeta^2)/2(\omega-\zeta)\}$.
Hence, the BCS logarithm is generated for energies \mbox{$\omega \sim \sqrt{m_g^2+\zeta^2}+\Lambda_y$} 
with $0\leq y<1$, since then $\xi_{\rm max}\sim \Lambda_y $. For such energies we have
\bea\label{tpole}
{\cal A}^{t}_{\rm pole}(\omega,\vk)
&\sim&g^2\int\limits_0^{\xi_{\rm max}}\frac{d\xi}{\epsilon_\vq}\,{\rm Re}\,\phi(\epsilon_\vq,\vq)
\sim g\,\phi\,\cos\left( \frac{\pi\,y}{2}\right)\;.
\eea
For larger energies up to $2\mu$ one has ${\cal A}^{t}_{\rm pole}(\omega,\vk) \sim g\,\phi$.  

In the longitudinal gluon sector we approximate 
\bea \label{rholapp}
\rho^\ell_{\rm pole}(\omega_\ell(\vp), \vp) \simeq -\frac{\omega_\ell(\vp)}{2p^2}
\eea
for gluon momenta $p$ not much larger than  $m_g$. Such values of $p$ are guaranteed
 if we consider energies of the form \mbox{$\omega = \sqrt{m_g^2+\zeta^2} +\Lambda_{y_1}$}
  with $\bar g < y_1 < 1$.  
As in the transversal case we simplify $\omega_{\ell}(\vp)\simeq \sqrt{p^2+m_g^2}$ and find
\bea\label{lpole}
{\cal A}^{\ell}_{\rm pole}(\omega,\vk)&\sim&
g^2\int\limits_0^{\xi_{\rm max}}\frac{d\xi}{\epsilon_\vq}\,{\rm Re}\,\phi(\epsilon_\vq,\vq)
\!\!\!\!\int\limits_{\sqrt{m_g^2+|\xi-\zeta|^2}}^{2\mu}\!\!\!\!\!\! d\omega_\ell\,
\frac{\omega_\ell^2}{\omega^2_\ell-m_g^2}
\,\delta(\omega^*-\omega_{\ell})\;,
\eea
where $\xi_{\rm max}$ is defined as in the transversal case. For the 
considered energies one has $\xi_{\rm max}\sim \Lambda_{y_1}$. We find
\bea\label{l4}
{\cal A}^{\ell}_{\rm pole}(\omega,\vk)
&\sim& g^2 \int\limits_0^{\xi_{\rm max}}\frac{d\xi}{\epsilon_\vq}\,{\rm Re}\,\phi(\epsilon_\vq,\vq)
\frac{(\omega-\epsilon_\vq)^2}{(\omega-\epsilon_\vq)^2-m_g^2}
\non
&\sim&  g\,\phi\,\frac{\omega}{\omega-m_g}\,\cos\left( \frac{\pi\,y_1}{2}\right)\;,
\eea
where $\xi_{\rm max}\ll \omega$ and $\omega \agt m_g$ was exploited in order to simplify
the fraction under the integral, $\omega^2/(\omega^2-m_g^2)\sim\omega/(\omega-m_g)$.
Furthermore, the BCS logarithm has cancelled one power of $g$.
Hence, for $\zeta \sim m_g$ one has ${\cal A}^{\ell}_{\rm pole} \sim g\,\phi$. For quarks
exponentially close to the Fermi surface with $\zeta \sim \Lambda_{y_2/2}$ and
$y_2> 2\,\bar g$, we find
\bea\label{l3}
{\cal A}^{\ell}_{\rm pole}(\omega,\vk) \sim g\,\phi
\,\left(\frac{M}{\phi}\right)^y\cos\left( \frac{\pi\,y_1}{2}\right)\;,
\eea
where $y\equiv {\rm min}\{y_1,y_2\}$. 
For much larger energies, $\omega \gg m_g$, we have to consider 
gluon momenta $p \gg m_g$, for which the spectral density of longitudinal gluons  
is exponentially suppressed
\bea\label{rholpoleexp}
\rho_{\rm pole}^{\ell}(\omega_l(\vp), \vp) \sim \frac{\exp\left(-\frac{2p^2}{3m_g^2} \right)}{p}\;.
\eea
We find for  $m_g \ll \omega<2\mu$ with $\omega_l(\vp) \simeq p$
\bea\label{lpole2}
{\cal A}^{\ell}_{\rm pole}(\omega,\vk)&\sim&g^2\int\limits_0^{\Lambda_{\rm q}}
\frac{d\xi}{\epsilon_\vq}\,{\rm Re}\,\phi(\epsilon_\vq,\vq)
\!\!\!\!\int\limits_{\sqrt{m_g^2+|\xi-\zeta|^2}}^{2\mu}\!\!\!\!\!\! 
 d\omega_\ell\,\exp\left(-\frac{2\omega_\ell^2}{3m_g^2} \right)
\,\delta(\omega^*-\omega_{\ell})\non
&\sim& g^2\phi\int\limits_{\Lambda_1}^{\Lambda_0}\frac{d\xi}{\xi}\,
\exp\left[-\frac{2(\omega-\xi)^2}{3m_g^2} \right]
\sim g\,\phi \,\exp\left(-\frac{2\omega^2}{3m_g^2} \right)\;,
\eea
which is the continuation of Eq.\ (\ref{l4}) to large energies and for all $\zeta \leq \Lambda_{\rm q}$. 
The results for ${\cal A}^{\ell,t}_{\rm pole}$ are summarized in Table \ref{tableApolezeta<M}.
\begin{table}  
\centerline{\begin{tabular}[t]{|c||c|c|c|c|}
\hline
 &~  $\omega<\sqrt{m_g^2+\zeta^2}~ $
  &~  $\omega \sim \sqrt{ m_g^2+\zeta^2}+\Lambda_{1<y_1<\bar g}$
 ~ & ~$m_g\ll\omega<2\mu$~
\\ \hline\hline
~ ${\cal A}_{\rm pole}^t$ ~  &  0 & $g\,\phi$ & $g\,\phi$
\\ \hline
${\cal A}_{\rm pole}^\ell$  & 0 	& $g\,\phi\left(\frac{M}{\phi}\right)^y
\cos\left( \frac{\pi\,y_1}{2}\right)$  & 
~ $g\,\phi \,\exp\left(-\frac{2\omega^2}{3m_g^2} \right)$~ 
\\ \hline
\end{tabular}}
\caption{Estimates for ${\cal A}_{\rm pole}^{\ell,t}$  at different 
 energy scales and for $\zeta \sim \Lambda_{y_2}$, where $0\leq y_2\leq 1$ and
  $y\equiv {\rm min}\{y_1,y_2\}$. }
\label{tableApolezeta<M}
\end{table}
In the following Sec.\ \ref{estB} these results will be inserted into Eq.\ (\ref{ImMT0}) for 
$\rho_\phi$ in order to estimate $\cal B$.

\subsection{Estimating ${\cal B}$} \label{estB}  

The term  ${\cal B}$ in Eq.\ (\ref{AB}) can be estimated using
\bea\label{Blt}
{\cal B}^{\ell,t}_{\rm cut}(\omega,\vk) \sim g^2\int\limits_0^{\Lambda_{\rm q}} \frac{d\xi}
{\epsilon_\vq} \int\limits _0^\omega dq_0 \sum\limits_{\sigma=\pm}
\frac{\sigma\,{\cal A}(q_0 ,\vq)}{q_0-\sigma \epsilon_\vq} \int\limits _\lambda^{\Lambda_{\rm gl}}
 dp\,p \,\rho^{\ell,t}_{\rm cut}(\omega^\prime,\vp)\;,
\eea
for the Landau-damped gluon sector. We introduced 
$\omega^\prime\equiv\omega-q_0<\omega$ and 
$\lambda\equiv {\max}(|\xi -\zeta|,\omega^\prime)$. Furthermore, we set $Z^2(\omega)\sim 1$.
From the condition
$\lambda<\Lambda_{\rm gl}$ in Eq.\ (\ref{Blt}) and from ${\cal A}(q_0,\vq) = 0$ for
$q_0 > 2\mu$ it follows that ${\cal B}^{\ell,t}_{\rm cut}(\omega,\vk) = 0$ 
for $\omega>\Lambda_{\rm gl}+2\mu\sim 3\mu$.  Inserting the approximative forms (\ref{appcut}) for 
$\rho^{\ell,t}_{\rm cut}$ into Eq.\ (\ref{Blt}) the integration over $p$ can be performed analogously to
Eqs.\ (\ref{t},\ref{pintlong}). In the transverse case one finds
\bea\label{t2}
\hspace*{-0.6cm}
{\cal B}^t_{\rm cut}(\omega,\vk)\sim
 g^2\int\limits_0^{\Lambda_{\rm q}} \frac{d\xi}{\epsilon_\vq} \int\limits _0^\omega dq_0 
 \sum\limits_{\sigma=\pm}\frac{\sigma\,{\cal A}(q_0,\vq) }{q_0-\sigma \epsilon_\vq}
 \left[\arctan\left(\frac{\Lambda_{\rm gl}^3}{M^2\omega^\prime}\right)-
 \arctan\left(\frac{\lambda^3}{M^2\omega^\prime}\right)\right]\;.
\eea
Analogously to  Eq.\ (\ref{t}), one first determines the domains of $\omega,\,\zeta$, and 
$\xi$, where the arctangents in the squared brackets do not cancel.
Since $\omega^\prime<\omega <3\mu$ it is $\Lambda_{\rm gl}^3/M^2\omega^\prime \gg 1$
 and the first arctangent in the squared brackets may be set equal to $\pi/2$. 
Furthermore, one finds that the argument of the second arctangent is not very large as
 long as the conditions $\omega - M \lesssim q_0$ and 
 $|\xi-\zeta|^3\lesssim M^2(\omega-q_0)$ are
fulfilled. In order to satisfy the first condition, we restrict the integral over $q_0$ to the region
  ${\rm max}\{0,\omega - M\}< q_0 <\omega$.
The second condition becomes less restrictive if simplified to $|\xi-\zeta|^3\lesssim M^2\omega$. 
The resulting  estimate for ${\cal B}^{t}_{\rm cut}$ will turn out to be small so that a
more elaborate estimate is not necessary.
For energies $\omega\sim \phi$ one may use Eq.\ (\ref{Aphi}) to estimate 
${\cal A} \sim{\cal A}^t_{\rm cut}\sim g^2\phi$. For $\zeta \ll M$ one has
\bea
{\cal B}^t_{\rm cut}(\phi,\vk) \sim
g^4 \phi \int\limits_0^{\Lambda_{1/3}} \frac{d\xi}{\epsilon_\vq}\,
\ln\left| \frac{\epsilon_\vq-\phi}{\epsilon_\vq + \phi}\right| \sim g^4\phi\;.\label{Bcutphi}
\eea
The logarithm under the integral prevents the generation of the BCS logarithm, cf.\ Appendix
\ref{nonBCS}. For $\zeta \lesssim M$ the integration over $\xi$ is restricted to the region 
$|\xi-\zeta|< \Lambda_{1/3}$. This yields the estimate 
${\cal B}^t_{\rm cut}(\phi,\vk)\sim g^4\phi \,(\Lambda_{1/3}/M)
 \sim g^4\phi\, (\phi/M)^{1/3}$.

For $\omega\sim \Lambda_y$ with $ 0 \leq y < 1$ we conservatively estimate
${\cal A }\sim {\cal A}^t_{\rm cut} \sim g\,\phi$, cf.\ Eq.\ (\ref{Aphi2}), and obtain 
similarly to Eq.\ (\ref{Bcutphi})
\bea\label{est1}
{\cal B}^t_{\rm cut}(\Lambda_y,\vk)\sim
g^3 \phi \int\limits_0^{\Lambda_{y/3}} \frac{d\xi}{\epsilon_\vq}\,\ln\left|
\frac{\epsilon_\vq-\Lambda_y}{\epsilon_\vq + \Lambda_y}\right| \sim g^3\phi\;,
\eea
where we assumed $\zeta \ll M$. Again no BCS logarithm was generated due to the additional 
logarithm. For $\zeta \lesssim M$ the integration over $\xi$ is restricted to the region 
$|\xi-\zeta|< \Lambda_{y/3}$. If we conservatively estimate ${\cal A }\sim g\,\phi$ 
throughout this region, we obtain 
${\cal B}^t_{\rm cut}(\phi,\vk)\sim g^3\phi\, (\phi/M)^{y/3}$.

For energies $\omega \sim M$ and larger, the condition $|\xi-\zeta|^3\lesssim M^2\omega$
is fulfilled for all $\xi < \Lambda_{\rm q}$. 
Considering $\omega \sim m_g +\Lambda_y$ with $0\leq y <1$ we find that the dominant
 contribution comes from ${\cal A} \sim {\cal A}^\ell_{\rm pole}$, cf.\ Eq.\ (\ref{l3}), 
when integrating over $m_g+\Lambda_1<q_0<m_g+\Lambda_y$
\bea\label{est2}
{\cal B}^t_{\rm cut}(m_g+\Lambda_y,\vk)&\sim& 
g^3 \phi \int\limits_0^{\Lambda_{\rm q}} \frac{d\xi}{\epsilon_\vq} 
\int\limits _{m_g+\Lambda_1}^{m_g+\Lambda_y} dq_0 \sum\limits_{\sigma=\pm}
\frac{\sigma}{q_0-\sigma \epsilon_\vq}\,\frac{q_0}{q_0-m_g}\non
&\sim &
g^2 \phi \int\limits_0^{\Lambda_{\rm q}} \frac{d\xi}{\epsilon_\vq}\,\sum
\limits_{\sigma=\pm}\frac{\sigma\,m_g}{\sigma \epsilon_\vq-m_g}\int\limits _1^y dy^\prime
\sim g^2 \phi \int\limits_0^{\Lambda_{\rm q}} d\xi\,\frac{m_g}{\xi^2-m_g^2}\non
&\sim& g^2\,\phi\;.
\eea
The contributions from the other gluon sectors are estimated with
${\cal A} \sim g\,\phi$, cf.\ Eqs.\ (\ref{Acuthigh},\ref{lpole}),
\bea\label{est4}
g^3 \phi \int\limits_0^{\Lambda_{\rm q}} \frac{d\xi}{\epsilon_\vq} 
\int\limits _{\omega-M}^{\omega} dq_0\,
\sum\limits_{\sigma=\pm}\frac{\sigma}{q_0-\sigma \epsilon_\vq}&\sim&
g^3 \phi \int\limits_0^{\Lambda_{\rm q}} \frac{d\xi}{\epsilon_\vq} \left(
\ln\left|\frac{\omega-\epsilon_q}{\epsilon_q+\omega} \right|
-\ln\left|\frac{\omega-M+\epsilon_q}{\omega-M-\epsilon_q} \right| \right)\non
&\sim& g^3 \phi \left( \frac{M}{\omega}\right)^2\;,
\eea
where the logarithms again prevent the generation of the BCS logarithm. The factor 
$(M/\omega)^2$ arises from expanding the logarithms for $\omega \gg M$. 
Hence, for energies of the form $\omega \sim m_g +\Lambda_y$
 with $0\leq y <1$ we have
\bea
\label{est3}
{\cal B}^t_{\rm cut}(\omega,\vk)&\sim& g^2\phi\;,
\eea
while for larger energies  ${\cal A}^\ell_{\rm pole}$  does  not contribute anymore
and ${\cal B}^t_{\rm cut}(\omega,\vk) \sim g^3\phi\,(M/\omega)^2$, 
cf.\  Eq.\ (\ref{est4}).

For the cut of the longitudinal gluons one obtains
\bea\label{l2}
{\cal B}^\ell_{\rm cut}(\omega,\vk)&\sim&
 g^2\int\limits_0^{\Lambda_{\rm q}} \frac{d\xi}{\epsilon_\vq} \int\limits _0^\omega 
 dq_0 \sum\limits_{\sigma=\pm}\frac{\sigma\,{\cal A}(q_0,\vq) }
 {q_0-\sigma \epsilon_\vq}\;\omega^\prime\;{\cal I}(\lambda)\;,
\eea
where ${\cal I}(\lambda)$ is defined in Eq.\ (\ref{pintlong}).
Analogously to the analysis of ${\cal A}_{\rm cut}^\ell$ one finds for
 $\omega \sim \phi$
\bea
{\cal B}^\ell_{\rm cut}(\phi,\vk)&\sim& g^4\phi\int\limits_0^{\Lambda_{\rm q}}
 \frac{d\xi}{\epsilon_\vq} \int\limits _0^\phi dq_0 \sum\limits_{\sigma=\pm}
 \frac{\sigma}{q_0-\sigma \epsilon_\vq}\,\frac{\phi}{M} 
\sim
 g^4\phi\,\frac{\phi}{M} \int\limits_0^{\Lambda_{\rm q}} \frac{d\xi}{\epsilon_\vq}
  \,\ln\left| \frac{\epsilon_\vq-\phi}{\epsilon_\vq + \phi}\right|\non
&\sim&
 g^4\phi\,\frac{\phi}{M}\;, \label{prevent2}
\eea
and similarly for $\omega\sim \Lambda_y$ with $ 0\leq y < 1$
\bea\label{prevent3}
{\cal B}^\ell_{\rm cut}(\Lambda_y,\vk)\sim
g^3\phi\left(\frac{\phi}{M}\right)^{y}\;.
\eea
For $\omega \sim m_g +\Lambda_y$ with $0\leq y < 1$ we can simplify 
 $\omega^\prime\, {\cal I}(\lambda)\simeq1$ and find as in the transversal case, cf.\ 
 Eq.\ (\ref{est2}), 
\bea
{\cal B}^\ell_{\rm cut}(m_g+\Lambda_y,\vk)&\sim& g^2 \phi \;.
\eea
In the limit of large energies, $M\ll \omega \sim \Lambda_{\rm gl}\sim \mu$, we estimate the 
integral over the range $\omega-\Lambda_{\rm gl}<q_0 < \omega$ by substituting 
${\cal A} \sim g\,\phi$ and obtain with $\omega^\prime\, {\cal I}(\lambda) \sim 1$
\bea\label{large1}
{\cal B}^\ell_{\rm cut}(\omega,\vk)&\sim&
g^3 \phi \int\limits_0^{\Lambda_{\rm q}} \frac{d\xi}{\epsilon_\vq} 
\int\limits _{\omega-\Lambda_{\rm gl}}^{\omega} dq_0 \sum\limits_{\sigma=\pm}
\frac{\sigma}{q_0-\sigma \epsilon_\vq} 
\sim g^3 \phi \int\limits_0^{\Lambda_{\rm q}} d\xi 
\int\limits _{\omega-\Lambda_{\rm gl}}^{\omega} \frac{dq_0}{q_0^2} \non
&\sim& g^3\phi \,\frac{M\Lambda_{\rm gl}}{\omega^2}\sim  g^3\phi\, \frac{M}{\omega}\;.
\eea
Hence, also ${\cal B}^\ell_{\rm cut}$ becomes small in the limit of large energies.
The estimates for ${\cal B}^{\ell,t}_{\rm cut}$ are summarized in 
Tables \ref{tableBcutzeta<M}  and \ref{tableBcutzetasimM}.

\begin{table}  
\centerline{\begin{tabular}[t]{|c||c|c|c|c|}
\hline
  &~  $\omega\sim\phi $~ &~  $\omega\sim\Lambda_{1>y>0}$~ &
   ~$\omega\sim m_g+\Lambda_{1>y>0}$~ & ~$m_g\ll\omega<3\mu$~
\\ \hline\hline
~ ${\cal B}_{\rm cut}^t$ ~  &  $g^4\phi$ & $g^3\phi$ & $g^2\phi$ & 
$g^3\phi\left(\frac{M}{\omega} \right)^2$
\\ \hline
${\cal B}_{\rm cut}^\ell$  & ~$g^4\phi\,\frac{\phi}{M}$~ & ~
$g^3\phi\left(\frac{\phi}{M}\right)^y$ ~ &  $g^2\phi$  & ~ $g^3\phi\,\frac{M}{\omega} $~ 
\\ \hline
\end{tabular}}
\caption{Estimates for ${\cal B}_{\rm cut}^{\ell,t}$ at different energy
scales and $\zeta\ll M$.}
\label{tableBcutzeta<M}
\vspace{0.5cm} 
\centerline{\begin{tabular}[t]{|c||c|c|c|c|}
\hline
  &~  $\omega\sim\phi $~ &~  $\omega\sim\Lambda_{1>y>0}$~ &
   ~$\omega\sim m_g+\Lambda_{1>y>0}$~ & ~$m_g\ll\omega<3\mu$~
\\ \hline\hline
~ ${\cal B}_{\rm cut}^t$ ~  &  $~g^4\phi\,\left(\frac{\phi}{M}\right)^{1/3}~$ &
 $~g^3\phi\,\left(\frac{\phi}{M}\right)^{y/3}$ & 
$g^2\phi$ & $g^3\phi\left(\frac{M}{\omega} \right)^2$
\\ \hline
${\cal B}_{\rm cut}^\ell$  & ~$g^4\phi\,\frac{\phi}{M}$~ & ~
$g^3\phi\left(\frac{\phi}{M}\right)^y$ ~ &  $g^2\phi$  & ~ $g^3\phi\,\frac{M}{\omega} $~ 
\\ \hline
\end{tabular}}
\caption{Estimates for ${\cal B}_{\rm cut}^{\ell,t}$ at different energy scales
  and $\zeta\lesssim M$.}
\label{tableBcutzetasimM}
\end{table}

In the undamped gluon sector the term  ${\cal M}^{\ell,t}_{{\cal B},T=0}$ in Eq.\ (\ref{ImMT0}) 
gives the contribution 
\bea\label{Bpole}
{\cal B}^{\ell,t}_{\rm pole}(\omega,\vk) \sim g^2\int\limits_0^{\Lambda_{\rm q}} \frac{d\xi}
{\epsilon_\vq} \int\limits _0^\omega dq_0 \sum\limits_{\sigma=\pm}\frac{\sigma\,
\rho_\phi(q_0,\vq)}{q_0-\sigma \epsilon_\vq} \int\limits _{|\zeta-\xi|}^{2\mu} dp\,p \,
\rho^{\ell,t}_{\rm pole}(\omega^\prime,\vp)\,\delta[\omega^\prime-\omega_{\ell,t}(\vp)]\;.
\eea
Due to the restriction $p< 2\mu$ it follows with similar arguments as for
 ${\cal B}^{\ell,t}_{\rm cut}$ that ${\cal B}^{\ell,t}_{\rm pole}(\omega,\vk)=0$ for $\omega >4\mu$. 
For the transversal sector we employ analogous approximations as for 
${\cal A}^{t}_{\rm pole}$ and obtain
\bea
{\cal B}^{t}_{\rm pole}(\omega,\vk)&\sim& g^2\int\limits_{\Lambda_1}^{\Lambda_0} \frac{d\xi}{\xi} 
\int\limits _0^\omega dq_0 \sum\limits_{\sigma=\pm}
\frac{\sigma\,{\cal A}(q_0,\vq)}{q_0-\sigma \epsilon_\vq} 
\int\limits _{\sqrt{m_g^2+|\xi-\zeta|^2}}^{2\mu}\!\!\! d\omega_t \,\delta(\omega^\prime-\omega_{t})\;.
\eea
This contribution is non-zero only if $\omega>m_g$. First we consider energies 
$\omega \sim m_g + \Lambda_{2y_1}$ and $\zeta \sim \Lambda_{y_2}$, and analyze the
two cases $y_1 <y_2$ and $y_1 >y_2$ separately. In the first case, the condition 
$\omega>\sqrt{m_g^2+|\xi-\zeta|^2}$ requires $0<\xi<\Lambda_{y_1}$ and consequently
\bea
{\cal B}^{t}_{\rm pole}(\omega,\vk)&\sim& 
g^3\phi\int\limits_{\Lambda_1}^{\Lambda_{y_1}} \frac{d\xi}{\xi} \,
\,\ln\left|\frac{\omega-\sqrt{m_g^2+\xi^2}-\xi}{\omega-\sqrt{m_g^2+\xi^2}+\xi}\right|
\sim g^3\phi\;,\label{Bcuttrans1}
\eea
where ${\cal A}\sim g\,\phi$ and the logarithm prevents the BCS logarithm. The second case,  
$y_1 >y_2$, leads to the condition 
\mbox{$ \Lambda_{y_2} -\Lambda_{y_1} <\xi < \Lambda_{y_2} + \Lambda_{y_1}$}
and we find
\bea
{\cal B}^{t}_{\rm pole}(\omega,\vk)&\sim& 
g^3\phi\int\limits_{\Lambda_{y_2}-\Lambda_{y_1}}^{\Lambda_{y_2}+\Lambda_{y_1}} 
\frac{d\xi}{\xi} \,\
\ln\left|\frac{m_g-\sqrt{m_g^2+\Lambda_{y_2}^2} -\Lambda_{y_2}}
{m_g-\sqrt{m_g^2+\Lambda_{y_2}^2} +\Lambda_{y_2}}\right|
\sim g^3\phi\,\left( \frac{\phi}{M}\right)^{y_1}\;,\label{Bcuttrans2}
\eea
where in the last step the logarithm was estimated to be of order $(\phi/M)^{y_2}$ and the 
integral over $\xi$ to be of order  $(\phi/M)^{y_1-y_2}$.

For $\omega > 2m_g$ the upper boundary for the integral over $\xi$ is given by 
$\Lambda_{\rm q}$ without  further restrictions. The integral over $q_0$ runs over values 
$q_0 >m_g$ and therefore receives contributions from ${\cal A}^\ell_{\rm pole}$, cf.\ Eq.\ (\ref{l3}). 
As a consequence one finds 
\bea\label{est8}
{\cal B}^{t}_{\rm pole}(\omega,\vk)\sim g^2\phi
\eea
 analogously to Eq.\ (\ref{est2}). For $\omega \gg m_g$ the additional contributions from 
 $2m_q<q_0 <\omega$ are only $\sim g^3 \phi$, as can be seen in the same way as in
 Eq.\ (\ref{est4}), and therefore ${\cal B}^{t}_{\rm pole}\sim g^2\phi$. However, for 
 $\omega\agt 2\mu+2m_g$ the condition $\omega^\prime = \omega_\ell$ can be fulfilled only
  for $q_0>\omega - 2\mu\agt 2m_g $, where ${\cal A }\sim g\,\phi$, and one finds
\bea\label{large2}
{\cal B}^{t}_{\rm pole}(\omega,\vk)&\sim& 
g^3\phi\int\limits_{\Lambda_1}^{\Lambda_0} \frac{d\xi}{\xi} \int\limits _{\omega-2\mu}^\omega dq_0
 \sum\limits_{\sigma=\pm}\frac{\sigma}{q_0-\sigma \epsilon_\vq}
\sim g^3\phi\int\limits_{\Lambda_1}^{\Lambda_0} d\xi \int\limits _{\omega-2\mu}^\omega 
\frac{dq_0}{q_0^2}\non
&\sim& g^3\phi\,\frac{M\,2\mu}{\omega^2}\sim g^3\phi\,\frac{M}{\omega}\;.
\eea

In the longitudinal sector the analysis starts similarly with energies 
$\omega  \sim m_g +\Lambda_{2y_1}$ and $\zeta \sim \Lambda_{y_2}$. Since this restricts
 the gluon momentum to $p \lesssim m_g$, we apply Eq.\ (\ref{rholapp}) and obtain
\bea
{\cal B}^{\ell}_{\rm pole}(\omega,\vk)
&\sim& g^2\int\limits_{\Lambda_1}^{\Lambda_0} \frac{d\xi}{\xi}\int\limits _0^{\omega} dq_0 
\sum\limits_{\sigma=\pm}\frac{\sigma\,{\cal A}(q_0,\vq) }{q_0-\sigma \epsilon_\vq}
\int\limits _{\sqrt{m_g^2+|\xi-\zeta|^2}}^{\omega} \!\!\!\!\!d\omega_\ell\,\frac{\omega_\ell^2}
{\omega^2_\ell-m_g^2} 
\,\delta(\omega^\prime-\omega_{\ell})\;.
\eea
In the case that $y_1< y_2$, the condition $\omega>\sqrt{m_g^2+|\xi-\zeta|^2}$ requires
$0<\xi<\Lambda_{y_1}$ and one finds
\bea
{\cal B}^{\ell}_{\rm pole}(\omega,\vk)
&\sim& g^2\int\limits_{\Lambda_1}^{\Lambda_{y_1}} \frac{d\xi}{\xi}\int\limits _0
^{\omega-\sqrt{m_g^2-|\xi-\zeta|^2}} \!\!\!\!\!dq_0\sum\limits_{\sigma=\pm}\frac{\sigma\,
{\cal A}(q_0,\vq)}{q_0-\sigma \epsilon_\vq}\;\frac{(\omega-q_0)^2}{(\omega-q_0)^2-m_g^2} \;.\label{est5}
\eea 
Since $q_0\leq\omega-\sqrt{m_g^2-|\xi-\zeta|^2}<\Lambda_{2y_1}$ is much smaller than 
$\omega \sim m_g+ \Lambda_{2y_1}$, we neglect $q_0$ against $\omega$ on the r.h.s\ of 
Eq.\ (\ref{est5}). Furthermore, we estimate ${\cal A}\sim g\,\phi$ and obtain similarly to
 Eq.\ (\ref{Bcuttrans1})
\bea
{\cal B}^{\ell}_{\rm pole}(\omega,\vk)
&\sim&  g^3\phi\,\frac{\omega}{\omega-m_g}\sim  g^3\phi\,\left(\frac{M}{\phi}\right)^{2y_1}
\;.\label{prevent4}
\eea
The case that $y_1> y_2$ leads to the condition
$\Lambda_{y_2} -\Lambda_{y_1} <\xi < \Lambda_{y_2} + \Lambda_{y_1}$, and we find 
similarly to Eq.\ (\ref{Bcuttrans2})
\bea
{\cal B}^{\ell}_{\rm pole}(\omega,\vk)
&\sim&  g^3\phi\,\left(\frac{\phi}{M} \right)^{y_1}\frac{\omega}{\omega-m_g}\sim
g^3\phi\,\left(\frac{M}{\phi}\right)^{y_1}
\;.
\eea

For larger energies $\omega \agt 2m_g$ the upper boundary of the integral over $q_0$ will 
just exceed $m_g$, where it is ${\cal A} \sim {\cal A}^\ell_{\rm pole}$, cf.\ Eq.\ (\ref{l3}). One 
finds analogously to ${\cal B}^t_{\rm pole}$, cf.\ Eq.\ (\ref{est8}),  that this gives the main 
contribution of order
\bea
{\cal B}^\ell_{\rm pole}(\omega,\vk)\sim g^2\phi \;.\label{est6}
\eea
In order to estimate ${\cal B}^{\ell}_{\rm pole}$ for energies $\omega> 2\mu+2m_g$, for which
 $2m_g<q_0<\omega$ and ${\cal A}$ is only  $\sim g\,\phi$, we have to employ 
Eq.\ (\ref{rholpoleexp}) 
\bea\label{lpole3}
{\cal B}^{\ell}_{\rm pole}(\omega,\vk)
&\sim& g^3\phi\int\limits_0^{\Lambda_{\rm q}}\frac{d\xi}{\epsilon_\vq}
\int\limits _{2m_g}^\omega dq_0 \sum\limits_{\sigma=\pm}
\frac{\sigma}{q_0-\sigma \epsilon_\vq}\int\limits_{m_g}^{2\mu}
 d\omega_\ell\,\exp\left(-\frac{2\omega_\ell^2}{3m_g^2} \right)
\,\delta(\omega^\prime-\omega_{\ell})\non
&&\sim g^3\phi\int\limits_{\Lambda_1}^{\Lambda_0}d\xi
\,\int\limits _{2m_g}^\omega \frac{dq_0}{q_0^2}\,
\exp\left[-\frac{2(\omega-q_0)^2}{3m_g^2} \right]
\sim g^3\phi\left(\frac{M}{\omega}\right)^2\;.
\eea
In the last step, the integral over $q_0$ was restricted to the region 
$\omega -m_g \lesssim q_0 <\omega$ due to the exponential function.
The estimates for ${\cal B}^{t,\ell}_{\rm pole}$ are summarized in 
Tables \ref{tableBpolezeta<M}  and \ref{tableBpolezetasimM}.

\begin{table}  
\centerline{\begin{tabular}[t]{|c||c|c|c|c|}
\hline
 & ~ $\omega < m_g+\Lambda_1$ ~ & ~$\omega \sim m_g+\Lambda_{1>y\geq0}$~ & 
 ~$2m_g<\omega <2\mu$~ 
& ~$2\mu<\omega<4\mu$~
\\ \hline\hline
~ ${\cal B}_{\rm pole}^t$ ~  &  0 & $g^3\phi$ & $g^2\phi$ & $g^3\phi\,\frac{M}{\omega}$
\\ \hline
${\cal B}_{\rm pole}^\ell$  & 0 	& ~$g^3\phi\left(\frac{M}{\phi}\right)^y$ ~ &  $g^2\phi$  & ~ 
$g^3\phi \,\left(\frac{M}{\omega}\right)^2$~ 
\\ \hline
\end{tabular}}
\caption{Estimates for ${\cal B}_{\rm pole}^{\ell,t}$ at different
energy scales and $\zeta\ll M$.}
\label{tableBpolezeta<M}
\vspace{1cm}
\centerline{\begin{tabular}[t]{|c||c|c|c|c|}
\hline
 & ~ $\omega < m_g+\Lambda_1$ ~ & ~$\omega \sim m_g+\Lambda_{1>y\geq0}$~ & 
 ~$2m_g<\omega <2\mu$~ 
& ~$2\mu<\omega<4\mu$~
\\ \hline\hline
~ ${\cal B}_{\rm pole}^t$ ~  &  0 & $g^3\phi\,\left(\frac{\phi}{M}\right)^{y/2}$ & $g^2\phi$ &
$g^3\phi\,\frac{M}{\omega}$
\\ \hline
${\cal B}_{\rm pole}^\ell$  & 0 	& ~$g^3\phi\left(\frac{M}{\phi}\right)^{y/2}$ ~ &  $g^2\phi$  & 
~ $g^3\phi \,\left(\frac{M}{\omega}\right)^2$~ 
\\ \hline
\end{tabular}}
\caption{Estimates for ${\cal B}_{\rm pole}^{\ell,t}$  at different energy
 scales and $\zeta\lesssim M$.}
\label{tableBpolezetasimM}
\end{table}
In the following the estimates for ${\cal A}$ and ${\cal B}$ are used to determine the order of 
magnitude of  ${\cal H}[{\cal A}]$ and ${\cal H}[{\cal B}]$.

\subsection{Estimating ${\cal H}[{\cal A}]$ and  ${\cal H}[{\cal B}]$ }\label{hilbert}

In order to determine the order of magnitude of  Re$\,\tilde \phi$ and the order of the 
corrections due to $\cal B$ we estimate the Hilbert transforms ${\cal H}[{\cal A}]$ and
${\cal H}[{\cal B}]$. For that the quark momentum has to be exponentially close to the Fermi surface, 
$\zeta \ll M$, because for quarks farther away from the Fermi surface Im$\,\phi$ cannot be
treated as a correction anymore. Furthermore, in that case the normal self-energy $\Sigma$ 
has to be accounted for self-consistently, which is beyond the scope of this work. 
The integral over $\omega$ in the Hilbert transforms ${\cal H}[{\cal A}]$ and ${\cal H}[{\cal B}]$ 
is split into the energy regimes which were used to estimate $\cal A$ and $\cal B$,
cf.\ Eq.\ (\ref{splitting}). As explained in the discussion after Eq.\ (\ref{splitting}) we select for each
energy regime the most dominant gluon sectors and estimate their  respective contributions to 
Re$\,\phi$ and Re$\,\tilde\phi$. 

At the smallest scale, $0\leq\omega \leq \Lambda_1$, one has 
Im$\,\phi \sim {\cal A}^t_{\rm cut} \sim g^2\phi$, cf.\ Eq. (\ref{Aphi}), which yields
\bea
&&\mathcal P\int\limits _0^{\Lambda_1} d\omega \,\sum\limits_{\sigma=\pm}
\frac{{\cal A}^t_{\rm cut}(\omega,\vk)}{\omega-\sigma\epsilon_\vk} \sim g^2 \phi\, 
\ln\left(\frac{\phi}{\epsilon_\vk}\right)\sim g^2\, \phi\;.
\eea
At the scale $\Lambda_1\leq \omega \leq \Lambda_{\bar g}$ one has
 Im$\,\phi \sim {\cal A}^t_{\rm cut} \sim g\,\phi$, cf. Eq.\ (\ref{Aphi2}). With the substitution 
 $d\omega/\omega = \ln(\phi/M)\,dy$ one finds
\bea\label{hil1}
&&\mathcal P\int\limits _{\Lambda_1}^{\Lambda_{\bar g}} d\omega \,\sum\limits_{\sigma=\pm}
\frac{{\cal A}^t_{\rm cut}(\omega,\vk)}{\omega-\sigma\epsilon_\vk}\sim g\,\phi
\int\limits _{\Lambda_1}^{\Lambda_{\bar g}}\frac{d\omega}{\omega}  \sim g\, 
\phi \ln\left(\frac{\phi}{M}\right)\int\limits _{1}^{{\bar g}}dy \sim \phi\;.\label{magcon}
\eea
The contribution from ${\cal A}^\ell_{\rm cut} \sim g\,\phi\, (\phi/M)^y$ at the same scale,
cf. Eq.\ (\ref{Aphi2b}),  can be shown to be much smaller,
\bea
&&\mathcal P\int\limits _{\Lambda_1}^{\Lambda_{\bar g}} d\omega \,
\sum\limits_{\sigma=\pm}\frac{{\cal A}^\ell_{\rm cut} (\omega,\vk)}{\omega-\sigma\epsilon_\vk}
\sim g\,\phi\,\ln\left(\frac{\phi}{M}\right)\int\limits _{1}^{\bar g} dy \,\left(\frac{\phi}{M}\right)^y  
\sim g\,\phi\,\frac{\phi}{M}\;.
\eea
At the scale $\Lambda_{\bar g}\leq \omega \leq \Lambda_0$ one has
 Im$\,\phi \sim {\cal A}^t_{\rm cut} \sim {\cal A}^\ell_{\rm cut} \sim g\,\phi$,
cf.\ Eq.\ (\ref{Aphi2}, \ref{Aphi2b}),  and finds
\bea\label{hil2}
&&\mathcal P\int\limits_{\Lambda_{\bar g}}^{\Lambda_0} d\omega \,\sum\limits_{\sigma=\pm}
\frac{{\cal A}^{\ell,t}_{\rm cut}(\omega,\vk)}{\omega-\sigma\epsilon_\vk}\sim g\,\phi
\int\limits_{\Lambda_{\bar g}} ^{\Lambda_0} \frac{d\omega}{\omega}  
\sim \phi\int\limits_{\bar g}^0 dy\sim g\,\phi\;.
\eea
For energies $\omega \agt m_g + \Lambda_y$ with $0\leq y < 1$ one has 
Im$\,\phi \sim {\cal A}^\ell_{\rm pole} \sim g\,\phi\, (M/\phi)^y$, cf. Eq.\ (\ref{l3}), and one finds with
 $d\omega = \ln(\phi/M) \,\Lambda_y\,dy$
\bea
\mathcal P\int\limits_{m_g+\Lambda_1}^{m_g+\Lambda_0} d\omega \,
\sum\limits_{\sigma=\pm}\frac{{\cal A}^\ell_{\rm pole}(\omega,\vk)}
{\omega-\sigma\epsilon_\vk}
&\sim&
 \frac{g\,\phi}{M}\ln\left(\frac{\phi}{M}\right)\int\limits_{1}^{0}  dy \,
 \Lambda_{y}\left(\frac{M}{\phi}\right)^y  
\sim\frac{\phi}{M}\int\limits_{1}^{0}  dy \,M  \sim \phi\;.\label{elcon}
\eea
For the regime $m_g < \omega < 2\mu$ we have 
Im$\,\phi \sim {\cal A}^t_{\rm pole}\sim g\,\phi$, 
cf.\ Eq.\ (\ref{tpole}), and obtain
\bea
\mathcal P\int\limits_{m_g}^{2\mu} d\omega \,\sum\limits_{\sigma=\pm}
\frac{{\cal A}^t_{\rm pole}(\omega,\vk)}{\omega-\sigma\epsilon_\vk}
\sim g\,\phi\int\limits_{m_g}^{2\mu} \frac{d\omega}{\omega} \sim g\,\phi 
\,\ln\left(\frac{\mu}{M} \right)\sim {g\,\phi}\;.\label{conlargeomega}
\eea
Finally, integrating over $2\mu < \omega < 4\mu$ with Im$\,\phi \sim
 {\cal B}^t_{\rm pole}\sim g^3\phi\,(M/\omega)$, cf.\ Eqs.\ (\ref{large1},\ref{large2}), one obtains
\bea
\mathcal P\int\limits_{2\mu}^{4\mu} d\omega \,\sum\limits_{\sigma=\pm}
\frac{{\cal B}^t_{\rm pole}(\omega,\vk)}{\omega-\sigma\epsilon_\vk}
\sim g^3\phi\,M\int\limits_{2\mu}^{4\mu} \frac{d\omega}{\omega^2} \sim 
g^3\phi \,\frac{M}{\mu} \sim g^4\phi\;.\label{converylargeomega}
\eea
From Eqs.\ (\ref{magcon}) and (\ref{elcon}) we conclude that Re$\,\tilde \phi\sim \phi$. 
Furthermore, ${\cal H}[{\cal B}]$ contributes to Re$\,\phi$ only at sub-subleading order. The 
corresponding corrections arise from the following sources. The first is ${\cal B}^t_{\rm cut}$,
which is $\sim g^2{\cal A}^t_{\rm cut}$  for $\omega \sim \Lambda_y$ with $\bar g<y<1$.
 After Hilbert transformation it yields a contribution of order $g^2\phi$ to Re$\,\tilde \phi$, 
cf.\ Eq.\ (\ref{hil1}), and is therefore of sub-subleading order. 
For $\omega \sim \Lambda_y$ with $0<y<\bar g$ we have 
${\cal B}^{\ell,t}_{\rm cut}\sim g{\cal A}^{\ell,t}_{\rm cut}$. From Eq.\ (\ref{hil2}) it follows that  
 ${\cal H}[{\cal B}^\ell_{\rm cut}]$ and ${\cal H}[{\cal B}^t_{\rm cut}]$ are of sub-subleading order. 
For $\omega = m_g+\Lambda_y$ with $0\leq y< $ one has
${\cal B}^\ell_{\rm pole}\sim g^2{\cal A}^\ell_{\rm pole}$ and a sub-subleading-order contribution 
seems possible, since the corresponding contribution from ${\cal A}^\ell_{\rm pole}$ is $\sim \phi$, 
cf.\ (\ref{elcon}). As the latter, however, combines with ${\hat\phi}$ to a subleading order term,
 cf.\ Sec.\ \ref{repro}, it would be interesting to investigate if also ${\cal B}^\ell_{\rm pole}$ 
 finds an analogous partner to cancel similarly. Moreover, we found that 
 ${\cal B}^{\ell,t}_{\rm pole}\sim g {\cal A}_{\rm pole}^t$ for $m_g<\omega < 2\mu$. From the
 estimate in Eq.\ (\ref{conlargeomega}) we conclude that the corresponding contributions 
 to Re$\,\phi$ are of  sub-subleading order. 
 The results are summarized in Table \ref{tablesummary}.
 In the next section it is analyzed at which order Im$\,\phi$ contributes to the local part of the gap function, 
 ${\hat\phi}$.

\subsection{The contribution of Im$\,\phi$ to ${\hat\phi}$}\label{phi0}

The gap equation for the energy-independent part ${\hat\phi}(\vk)$ is obtained by considering 
the integrals $I_0$ and $I_{k_0}$, cf.\ Eqs.\ (\ref{I0},\ref{Ik0}), in the limit 
$|k_0| \rightarrow \infty$. Since $p\lesssim 2\mu$, the gluon spectral densities 
$\rho^{\ell,t}(q_0,\vp)$ are nonzero only for $q_0 \lesssim 2\mu$. Consequently, 
the integral over $q_0$ in Eq.\ (\ref{Ik0}) is bounded by $-2\mu<q_0< 2\mu$. 
Then, due to the energy denominator under the integral, $I_{\rm k_0}$ tends to zero 
as $1/|k_0|$ for $|k_0| \rightarrow \infty$. In the second term on the r.h.s.\ of Eq.\ (\ref{I0})
one has ${\epsilon_\vq}<\Lambda_{\rm q} \sim g\mu$. It follows that for $k_0 \gg 2\mu>p$ the
 transverse  gluon propagator becomes $\Delta^t\sim 1/k_0^2$. In the longitudinal sector one has 
$\Delta^\ell \rightarrow -1/p^2$. Hence, in the limit $|k_0| \rightarrow \infty$ only  the longitudinal
contribution of the considered term does not vanish. Smilarly, one can argue that also in the first 
term on the r.h.s.\ of Eq.\ (\ref{I0}) only the contribution from the static electric gluon
 propagator remains. Consequently, we find for ${\hat\phi}(\vk)$
 \bea\label{localgapeq}
{\hat\phi}(\vk) &=& \frac{g^2}{3(2\pi)^2} \int\limits _{0}^{\Lambda_{\rm q}}\frac{d\xi}{\epsilon_\vq}
\int\limits _{|\xi-\zeta|}^{2\mu}\frac{dp}{p}\,{\rm Tr}_s^\ell (k,p,q)
\left[{\rm Re}\,\phi({\tilde \epsilon_\vq},\vq)\,
Z^2(\tilde\epsilon_\vq)\tanh\left(\frac{{\tilde \epsilon_\vq}}{2T} \right)\right.
\non&&
\left. +{\mathcal P}\int\limits _{-\infty}^\infty d\omega \,
\frac{\rho_{\phi}(\omega,\vq)}{{\tilde \epsilon_\vq}-\omega}\,Z^2(\omega)
\tanh\left(\frac{\omega}{2T} \right)\right]\;.
\eea
In the limit $T\rightarrow 0$ the hyperbolic functions simplify. After performing the integral over
 $p$ and with ${\rm Tr}_s^\ell (k,p,q)\sim 1$ and $Z^2(\omega)\sim1$ we obtain
\bea
{\hat\phi}(\vk) &\sim& g^2 \int\limits _{\Lambda_1}^{\Lambda_0}\frac{d\xi}{\xi}
\ln\left(\frac{2\mu}{|\xi-\zeta|}\right)
\left[ {\rm Re}\,\phi({\tilde \epsilon_\vq},\vq)
+{\mathcal P}\int\limits _{0}^\infty d\omega \,
\sum\limits_{\sigma=\pm}\,
\frac{\sigma\,\rho_{\phi}(\omega,\vq)}{\sigma{\tilde \epsilon_\vq}-\omega}\right]
\;,\label{phi02}
\eea
where the large logarithm arises from the $p-$integral. With that and assuming $\zeta \ll M$, the
integral containing ${\rm Re}\,\phi({\tilde \epsilon_\vq},\vq)$ is found to be of order $\phi$,
and hence ${\hat\phi}(\vk)\sim \phi$. The remaining
contribution from $\rho_\phi$ is identical to Eq.\ (\ref{splitting}) up to the extra $\sigma$ due
to the hyperbolic tangent. One can conservatively estimate this term by approximating 
$\rho_\phi \sim g\,\phi$ for $0<\omega<4\mu$ and all $\Lambda_1\leq\xi\leq\Lambda_0$, and
adding $\rho_\phi\sim g\,\phi\,(M/\phi)^y$ in the range $\omega\sim m_g+\Lambda_y\,,~ 1>y>0$. 
We find that the contributions from $\rho_\phi$ to ${\hat\phi}$ are of order $g^2\phi$ and hence 
of sub-subleading order.
This completes the proof that the contributions from Im$\,\phi$ to 
${\rm Re}\,\phi(\epsilon_\vk, \vk) = {\rm Re}\,\tilde\phi(\epsilon_\vk, \vk)+{\hat\phi}(\vk)$ 
are in total beyond subleading order.

\subsection{Re$\,\phi(\epsilon_\vk,\vk)$ to subleading order}\label{repro}

In the following we recover the real part of the gap equation to subleading order by Hilbert
transforming the imaginary part of the gap equation (\ref{Imphieq}) and adding the equation for the local 
gap, $\hat\phi(\vk)$, Eq.\ (\ref{localgapeq}). This shows how ${\hat\phi}\sim \phi$ and 
${\cal H}[{\cal A}_{\rm pole}^\ell]\sim \phi$ combine to a subleading-order contribution.
The gap equation for  Re$\,\tilde\phi(\epsilon_\vk,\vk)$ reads to subleading order
\bea
{\rm Re}\,\tilde\phi({ \epsilon_\vk},\vk)&=&
 -\frac{g^2}{3(2\pi)^2} \int\limits_0^{\Lambda_{\rm q}}d\xi
 \frac{Z^2(\tilde\epsilon_\vq)}{2\tilde\epsilon_\vq}\,{\rm Re}\,\phi({\tilde \epsilon_\vq},\vq)
 \,\tanh\left(\frac{\tilde\epsilon_\vq}{2T} \right)
\non
&&
\times\sum\limits_{\sigma=\pm}\left[
\int\limits_{|\xi-\zeta|}^{\Lambda_{\rm gl}}\!\!dp\,p\left\{{\rm Tr}_s^\ell (k,p,q)\left[\frac{1}{p^2}+
\Delta_{\rm HDL}^{\ell}(\epsilon_\vk-\sigma\tilde\epsilon_\vq,\vp)\right]+ {\rm Tr}_s^t (k,p,q)
\Delta^{t}_{\rm HDL}(\epsilon_\vk-\sigma\tilde\epsilon_\vq,\vp)\right\}\right.\non
&&\hspace{1.2cm}
+\left.
\int\limits_{\Lambda_{\rm gl}}^{2\mu}dp\,p\,{\rm Tr}_s^t (k,p,q)\,
 \Delta^{t}_{0,22}(\epsilon_\vk-\sigma\tilde\epsilon_\vq,\vp)
\right]\!,\label{rephitilde2}
\eea
where we used $\rho^\ell(\omega, \vp) \equiv 0$ for $p>\Lambda_{\rm gl}$ in the effective
 theory, cf.\ Eq.\ (\ref{hardspecl}). Furthermore, all terms $\sim \coth$ have been neglected. 
Adding Eq.\ (\ref{localgapeq}), the $1/p^2$-term from the soft electric gluon propagator in 
Eq.\ (\ref{rephitilde2}) restricts the $p-$integral of ${\hat\phi}$ from $\Lambda_{\rm gl}$ to 
$2\mu$. This is the aforementioned cancellation of ${\hat\phi}$ and 
${\cal H}[{\cal A}_{\rm pole}^\ell]$, which reduces these terms to the order $g\,\phi$.
After approximating the hard magnetic gluon propagator as 
$\Delta^{t}_{0,22}(\epsilon_\vk-\sigma\tilde\epsilon_\vq,\vp)= 
1/p^2 +O(\Lambda_{\rm q}/\Lambda_{\rm gl})$, one can combine it with the remaining 
contribution from ${\hat\phi}$. Using 
${\rm Tr}_s^\ell (k,p,q)-{\rm Tr}_s^t(k,p,q) =4 +O(\Lambda_{\rm q}/\Lambda_{\rm gl})$, 
one finally arrives at Eq.\ (124) of Ref.\ \cite{qwdhr}.

\subsection{${\rm Im}\,\phi(\epsilon_\vk+i\eta,\vk)$ exponentially close to the Fermi surface}\label{calcimphi}

In Sec.\ \ref{estA} and  \ref{estB} the contributions $\cal A$ and $\cal B$ to ${\rm Im}\,\phi$ have
been estimated for different regimes of $\omega$ and $\zeta$, cf.\ Tab.\ 
\ref{tableAcutzeta<M}-\ref{tableBpolezetasimM}. In the case that $\omega \sim \Lambda_y$
with $\bar g<y\leq 1$ and $\zeta < \Lambda_{y/3}$ we found 
${\cal A}^t_{\rm cut}$ to be the dominant contribution to ${\rm Im}\,\phi$. The cut of the
longitudinal gluons is suppressed by a factor $(\phi/M)^y$, while the gluon poles do not contribute 
at all. In other regions of $\omega$ and $\zeta$ different gluon sectors are shown to be dominant.
Furthermore, for $\omega >2m_g$ also the contributions from $\cal B$ 
would have to be considered, since there ${\cal B}$ is suppressed relative to ${\cal A}$ only by
one power of $g$ and therefore contributes at subleading order to
${\rm Im}\,\phi$. For $\omega \sim \Lambda_y$
with $\bar g<y \leq1$ and $\zeta < \Lambda_{y/3}$ we find for the imaginary part of the gap
\bea\label{Imphiexact}
{\rm Im}\,\phi(\omega+i\eta,\vk)
&\simeq&
\frac{g^2\,\pi}{3(2\pi)^2}\int\limits_{\Lambda_1}^{\omega}\frac{d\xi}{\xi}\,Z^2(\tilde\epsilon_\vq)\,
{\rm Re}\,\phi({\tilde \epsilon_\vq}\vq)\int\limits_\lambda^{\Lambda_{\rm gl}} dp
 \,\frac{2M^2\,\omega^*}{\pi}\,\frac{p^2}{p^6+(M^2\omega^*)^2}\non
&\simeq& \frac{g^2\,\pi}{9(2\pi)^2}\ln\left(\frac{\phi}{M}\right)\,\phi \int\limits_{1}^{y}
dy^\prime\,\sin\left(\frac{\pi\,y^\prime}{2} \right) \,\left(1-\frac{\bar g\,\pi\,y^\prime}{2} \right)
\non
&=&\bar{g}\,\phi \;\frac{\pi}{2}\cos\left(\frac{\pi\,y}{2} \right)+ {\cal O}(\bar g^2)\;,
\eea
where we substituted $\omega = \Lambda_y$ and $d\xi/\xi = dy^\prime\,\ln(\phi/M)$ and used 
$\ln(\phi/M)=-3\pi^2/(\sqrt{2}\,g)$. Furthermore, it was sufficient to approximate 
${\rm Tr}_s^t (k,p,q)\simeq -2$. This result agrees with Eq.\ (81) in Ref.\ \cite{ren} where
a different approach is used.

\section{Conclusions and Outlook}\label{outlook}

In this work we studied how the non-local nature of the gluonic interaction between quarks at high
densities affects the energy and momentum dependence of the (2SC) color-superconducting gap
function at weak coupling and zero temperature. For this purpose, energy and momentum have
been treated as independent variables in the gap equation.
By analytically continuing from imaginary to real energies and appropriately
choosing the contour of the integral over energies, we split the gap equation into two 
coupled equations: one for ${\rm Re}\, \phi$ and one for ${\rm Im}\, \phi$. 
 
In order to solve these equations self-consistently, the gap had to be estimated for all
energies and for all momenta satisfying $|k-\mu| \leq \Lambda_{\rm q}$, where 
$\Lambda_{\rm q}\sim g\mu$ is the quark cutoff of the effective theory 
employed in this work. For quarks exponentially close to the Fermi surface,
we have proven the previous conjecture that, to subleading order, 
one has $\phi = {\rm Re}\, \phi$, where ${\rm Re}\, \phi$ is the known subleading 
order solution for the real part of $\phi$, which neglects all contributions arising from the 
non-analyticities of $\phi$. 

Furthermore, we found that, exponentially close to the Fermi surface and for small energies,
only the cut of the magnetic gluon propagator contributes to Im$\,\phi$. Thus, the analytic solution 
of the imaginary part of the gap equation is rather simple, cf.\  Eq.\ (\ref{Imphiexact}). For energies of 
order $m_g$,  we showed that also the electric cut and the gluon poles contribute to Im$\,\phi$, 
cf.\ Tables \ref{tableAcutzeta<M}-\ref{tableBpolezetasimM}. The increase of the imaginary part 
with increasing energies can be interpreted as the opening of decay channels for the quasiquark 
excitations. The peak Im$\,\phi\sim g^2\mu$ occurring for energies just above $m_g$ reflects the
decay due to the emission of on-shell electric gluons.

Treating energy and momentum independently, the solution also includes Cooper pairs
further away from the Fermi surface, up to $|k-\mu|\sim g\mu$. This becomes important
when one is interested in extrapolating down to more realistic quark chemical potentials 
where the coupling  between the quarks becomes stronger: With increasing $g$ also quarks
away from the Fermi surface participate in Cooper pairing. These are not included in the
Eliashberg theory, where one assumes that Cooper pairing happens exclusively at the
Fermi surface. 

Finally, it would be interesting to generalize our analysis to non-zero temperatures. 
The dependence of $\phi(T)/\phi(T=0)$ on $T/T_c$, where $T_c$ is
the critical temperature for the onset of color superconductivity, agrees with
that of a weakly coupled BCS superconductor if one neglects Im$\,\phi$ \cite{rdpdhr}.
For strongly coupled superconductivity in metals
it is known \cite{schrieffer}, however, that Im$\,\phi$ is 
significantly modified at non-zero temperatures due to the presence of thermally 
excited quasiparticles.
 This in turn gives rise to important deviations from a BCS-like behavior of
$\phi(T)/\phi(T=0)$ at energies larger than the gap. An analogous analysis for color
superconductivity would be an interesting topic for future studies.

\section*{Acknowledgments}

I would like to thank Michael Forbes, Rob Pisarski, Hai-cang Ren, Dirk Rischke, Thomas Sch\"afer, Andreas Schmitt, 
Achim Schwenk, and Igor Shovkovy for interesting and helpful discussions. I thank the German 
Academic Exchange Service (DAAD) for financial support and the Nuclear Theory Group at the 
University of Washington for its hospitality.

\appendix

\section{Spectral representation of the gap function}\label{disper}

The function  $\phi(K)$ which solves the gap equation (\ref{gapequation1}) for imaginary
energies $k_0=i(2n+1)\pi T$, must be analytically continued towards the axis of real frequencies 
$k_0\rightarrow \omega + i\eta$ prior to any physical analysis. It is shown in the following that
the gap function must exhibit non-analyticities on the axis of real energies, in order to be 
a non-trivial function of energy. We require that the gap function converges
for infinite energies, i.e.,  $\phi(K) \rightarrow {\hat\phi}(\vk)$ for $|k_0|\rightarrow \infty$. 
Then $\phi(K)$ can be written as
\bea\label{tildehat}
\phi(K)\equiv \tilde\phi(K) +{\hat\phi}(\vk)\;,
\eea
where the energy dependence of $\phi(K)$ is contained in $ \tilde\phi(K)$ and $\tilde\phi(K) \rightarrow 0 $ 
for $|k_0|\rightarrow \infty$. The non-analyticities of $\phi(K)$ are contained in its spectral 
density
\bea \label{specphi3}
\rho_\phi(\omega,\vk) \equiv   \frac{1}{2\pi i}\,
\left[\phi(\omega+i\eta,\vk)-\phi(\omega-i\eta,\vk)\right]\;.
\eea
With Cauchy's theorem one obtains for any $k_0$ off the real axis
\bea\label{specphi}
\phi(k_0,\vk) = \int\limits_{-\infty}^\infty d\omega\,
\frac{\rho_\phi(\omega,\vk)}{\omega-k_0}+{\hat\phi}(\vk)\;,
\eea
where the first term is identified as $\tilde\phi(K)$ in Eq.\ (\ref{tildehat}). For 
 $k_0 = \omega +i\epsilon$ one obtains
\bea\label{phi3}
\phi(\omega +i\epsilon,\vk)
=\mathcal P\int\limits _{-\infty}^\infty d\omega^\prime \,
\frac{\rho_\phi(\omega^\prime,\vk)}{\omega^\prime -\omega}+
{\hat\phi}(\vk) + i\pi\rho_\phi(\omega,\vk)\;.
\eea
Hence, for real ${\hat\phi}(\vk)$ and $\rho_\phi(\omega,\vk)$
\bea\label{realphi2}
{\rm Re}\,\phi(\omega+i\epsilon,\vk)&=& \mathcal P\int\limits _{-\infty}^\infty d\omega^\prime \,
\frac{\rho_\phi(\omega^\prime,\vk)}{\omega -\omega^\prime}+{\hat\phi}(\vk)\;,\\
{\rm Im}\,\phi(\omega+i\epsilon,\vk)&=& {\pi}\,\rho_\phi(\omega,\vk)\;.\label{imphi2}
\eea
Furthermore, one finds the dispersion relations for $\tilde\phi(\omega+i\eta,\vk)$
\begin{subequations}
\bea\label{disp1}
{\rm Re}\,\tilde\phi(\omega+i\epsilon,\vk)&=& \frac{1}{\pi}\,\mathcal P\int\limits _{-\infty}^\infty 
d\omega^\prime \,\frac{{\rm Im}\,
\tilde\phi(\omega^\prime+i\epsilon,\vk)}{\omega^\prime -\omega}\;,\\
{\rm Im}\,\tilde\phi(\omega+i\epsilon,\vk)&=& -\frac{1}{\pi} \,\mathcal P\int\limits _{-\infty}^\infty
 d\omega^\prime \,\frac{{\rm Re}\,
\tilde\phi(\omega^\prime+i\epsilon,\vk)}{\omega^\prime -\omega}\;,\label{disp2}
\eea
\end{subequations}
i.e., $\rm{Im}\,\tilde\phi(\omega+i\epsilon,\vk)$ and $\rm{Re}\,\tilde\phi(\omega+i\epsilon,\vk)$
 are Hilbert transforms of each other, ${\rm Re}\,\tilde\phi= {\cal H}[{\rm Im}\,\tilde\phi]$.
It follows that $\tilde \phi \neq 0$ only if both Re$\, \tilde\phi$ and Im$\,\tilde\phi$ are nonzero.
 Consequently, the gap function $\phi$ is energy-dependent only if it has a nonzero imaginary part, 
Im$\,\phi \equiv {\rm Im}\,\tilde\phi\neq 0$, which in turn is  generated by its non-analyticities along 
the real $k_0$ axis, cf.\ Eqs.\ (\ref{specphi3}) and (\ref{imphi2}).

The energy dependence of the gap function $\phi(K)$ is constrained by the symmetry properties 
of the gap matrix $\Phi^+(K)$. It follows from the antisymmetry of the quark fields that the gap 
matrix must fulfill
\bea
C\,\Phi^+(K)\,C^{-1} =\left[  \Phi^+(-K)\right]^T\;,
\eea
cf.\ Eq.\ (B4) in  \cite{meanfield}. 
Since $C\gamma_5\Lambda^+_{\vk}C^{-1}= [\gamma_5\Lambda_{-\vk}^+]^T$ and
in the 2SC case $[J_3\tau_2]^T=J_3\tau_2$ it follows for the gap function
\bea 
\phi(K) &=& \phi(-K)\;.
\eea
Assuming that the gap function is symmetric under reflection of 3-momentum 
$\vk$, $\phi(k_0,\vk)=\phi(k_0,-\vk)$ one obtains with Eqs.\ (\ref{realphi2},\ref{imphi2})
\begin{subequations}
\bea
{\rm Re}\,\phi(\omega+i\eta,\vk)&=&{\rm Re}\,\phi(-\omega+i\eta,\vk)\;,\label{evenRe}\\
{\rm Im}\,\phi(\omega+i\eta,\vk)&=&-{\rm Im}\,\phi(-\omega+i\eta,\vk)\;,\label{oddIm}\\
\rho_{\phi}(\omega,\vk)&=&-\rho_{\phi}(-\omega,\vk)\;.\label{odd}
\eea
\end{subequations}
Consequently,  ${\rm Re}\,\phi$ is an even function of $\omega$ while 
${\rm Im}\,\phi$ and $\rho_\phi$ are odd. In the case of the color-flavor-locked
(CFL) phase the gap matrix has the structure $\Phi^+ \sim {\bm J}\cdot  {\bm I}$
in (fundamental) color and flavor space, where the matrices 
$(J_f)_{gh} \equiv -i\,\epsilon_{fgh}$ and $(I^k)^{mn} \equiv -i\,\epsilon^{kmn}$
act in color and flavor space, respectively \cite{DHRreview}. Since
$({\bm J}\cdot  {\bm I})^T={\bm J}\cdot  {\bm I}$, 
Eq.\ (\ref{evenRe}-\ref{odd}) are valid for the CFL phase, too.

Inserting Eq.\ (\ref{odd}) into Eq.\ (\ref{specphi}) one finds that $\phi(K)$ is real on the axis of 
imaginary energies, whereas $\phi(K)$ is complex for energies off the imaginary axis.

\section{Spectral representation of the gluon propagator}\label{specgluons}

In pure Coulomb gauge, the gluon propagator (\ref{splitgluonprop}) has the form
\be \label{propgen}
\Delta^{00}(P) = \Delta^{\ell}(P) \;\;\;\; , \;\;\;\;\;
\Delta^{0i}(P) = 0 \;\;\;\;, \;\;\;\;\;
\Delta^{ij}(P) =  (\delta^{ij} - \hat{p}^i \hat{p}^j) \,
\Delta^{t}(P)\;,
\ee
where $\Delta^{\ell,t}$ are the propagators for
longitudinal and transverse gluon degrees of freedom.
In the spectral representation one has \cite{specgluon}
\bea\label{longtransprops4}
\Delta^\ell(P) = - \frac{1}{p^2} + \int\limits_{-\infty}^{\infty} d \omega \, 
\frac{\rho^\ell(\omega,\vp)}{\omega-p_0}
 \,\,\,\,\,\,  ,  \,\,\,\,\,\,
\Delta^t(P) =  \int\limits_{-\infty}^{\infty} d \omega \, 
\frac{\rho^t(\omega,\vp)}{\omega-p_0}\;.
\eea
For hard gluons with momenta $p > \Lambda_{\rm gl}$ one finds
\begin{subequations}
\bea \label{hardspecl}
\rho_{0,22}^\ell(\omega,\vp) &=& 0\;,\\
\rho_{0,22}^t(\omega,\vp) &=& {\rm sign}(\omega)\delta(\omega^2-p^2)\;,\label{hardspect}
\eea
\end{subequations}
while for the soft, HDL-resummed gluons with  $p < \Lambda_{\rm gl}$ one has 
\cite{LeBellac,RDPphysicaA,specgluon}
\bea\label{specdens}
\!\!\!\!\!\!\!\!\!\!\!\!\!\!\! \rho_{\ell,t}(\omega,{\bf p}) & = & \rho^{\rm pole}_{\ell,t} 
(\omega,{\bf p})\, \left\{\delta \left[\omega - \omega_{\ell,t}({\bf p})\right] 
+\delta \left[\omega + \omega_{\ell,t}({\bf p})\right]\right\}
+\rho^{\rm cut}_{\ell,t}(\omega,{\bf p}) \, \theta(p-|\omega|)\;,
\eea
where
\begin{subequations} 
\begin{eqnarray}
\rho_\ell^{\rm pole}(\omega,{\bf p}) & = & \frac{\omega\, (\omega^2 - p^2)}{
p^2\,(p^2+3m_g^2-\omega^2)} \,\, , \\ \label{lcut}
\rho_\ell^{\rm cut}(\omega,{\bf p}) & = & \frac{2M^2}{\pi}\, \frac{\omega}{p}\,
\left\{ \left[ p^2+3\, m_g^2 \, \left( 1 -\frac{\omega}{2p} \, 
\ln \left| \frac{p+\omega}{p-\omega} \right| \right) \right]^2 + 
\left(2 M^2 \,\frac{\omega}{p} \right)^2 \right\}^{-1} \,\, ,\\
\rho_t^{\rm pole}(\omega,{\bf p}) & = & \frac{\omega\, (\omega^2 - p^2)}{
3m_g^2\,\omega^2-(\omega^2-p^2)^2} \,\, , \\ \label{tcut}
\rho_t^{\rm cut}(\omega,{\bf p}) & = & \frac{M^2}{\pi}\, \frac{\omega}{p}\,
\frac{p^2}{p^2-\omega^2}\,
\left\{ \left[ p^2+\frac{3}{2}\, m_g^2 \, 
\left( \frac{\omega^2}{p^2-\omega^2} +
\frac{\omega}{2p}\, \ln \left| \frac{p+\omega}{p-\omega} \right|
\right) \right]^2 +
\left(M^2\, \frac{\omega}{p} \right)^2 \right\}^{-1} \, \, .
\end{eqnarray}
\end{subequations}

\section{(Non-)Generating the BCS logarithm}\label{nonBCS}

The mixed scale $\Lambda_y$ introduced in Eq.\ (\ref{Lambday}) can be used to analyze the
generation of the BCS logarithm. For this purpose we have
split the integral over $\xi =k-\mu$ according to
\bea
g^2\int\limits_0^M \frac{d\xi}{\epsilon_\vq} \;\phi_q
=
g^2\int\limits_0^{\Lambda_1} \frac{d\xi}{\epsilon_\vq} \;\phi_q
+g^2\int\limits_{\Lambda_1}^{\Lambda_{\bar g}} \frac{d\xi}{\epsilon_\vq} \;\phi_q
+g^2\int\limits_{\Lambda_{\bar g}}^{\Lambda_{0 }} \frac{d\xi}{\epsilon_\vq} \;\phi_q
\eea
where we abbreviated $\phi_q \equiv {\rm Re}\,\phi(\epsilon_\vq, \vk)$. The first term 
is readily shown to be of order $g^2\phi$. The last term is of order $g^3\phi$, which is found after
 noting that $\phi_q \sim g\,\phi$, cf.\ Eq.\ (\ref{phileading}), and that $\Lambda_{\bar g}$ is smaller
 but of the order of $M$. The second term can be written as
\bea
g^2\int\limits_{\Lambda_{1}}^{\Lambda_{\bar g}} \frac{d\xi}{\epsilon_\vq} \;\phi_q &\sim&
g^2 \phi\, \ln\left( \frac{\phi}{M}\right) \int\limits_{{1}}^{{\bar g}} dy \;\sin\left( \frac{\pi\,y}{2}\right) 
\sim g\,\phi
\eea
where use was made of Eqs.\ (\ref{solution}-\ref{Lambday}) and the BCS logarithm $\ln(M/\phi)$
 being of order $1/g$. 
It is shown that the BCS logarithm arises from integrating over intermediate scales, 
$\Lambda_1 < \xi < \Lambda_{\bar g}$. This observation is useful for estimating
 numerous integrals. Note, 
however, that in the full QCD gap equation one also has the gluon propagator under the integral.
Then the region $\Lambda_1 <\xi< \Lambda_{\bar g}$ is additionally enhanced.

The contributions arising from $\rho_\phi(\omega)$ in the gap equation are suppressed,
 because the oddness of $\rho_\phi(\omega)$, cf.\ Eq.\ (\ref{odd}), prevents the generation 
 of the BCS logarithm. At many places, cf.\ Eqs.\ 
 (\ref{Bcutphi},\ref{est1},\ref{est4},\ref{prevent2},\ref{prevent3},\ref{Bcuttrans1},\ref{prevent4}),
  this oddness gives rise to logarithmic dependences of the following form 
\bea\label{noBCS1}
\int\limits_{\Lambda_1}^{\Lambda_0}\frac{d\xi}{\xi}\,
\ln\left|\frac{\xi + \Lambda_y}{\xi - \Lambda_y}\right|
=\int\limits_{\Lambda_1}^{\Lambda_y}\frac{d\xi}{\xi}\,
\ln\left(\frac{\xi + \Lambda_y}{\Lambda_y-\xi }\right)+ 
\int\limits_{\Lambda_y}^{\Lambda_0}\frac{d\xi}{\xi}\,
\ln\left(\frac{\xi + \Lambda_y}{\xi - \Lambda_y}\right)\;,
\eea
where $0\leq y \leq 1$. To show that the BCS logarithm is prevented we introduce the
dilogarithm \cite{Nielson}
\bea
{\rm Li}_2(x)\equiv \int\limits_x^0 \frac{d\xi}{\xi}\,\ln(1-\xi)\;,
\eea
which has the values
 $-\frac{1}{12} \pi^2 \equiv {\rm Li}_2\left(-1\right)\leq {\rm Li}_2\left(x\right)\leq{\rm Li}_2
 \left(1\right) \equiv  \frac{1}{6} \pi^2$ for $-1 \leq x\leq 1$. We write the first term on the r.h.s.\ 
 of Eq.\ (\ref{noBCS1}) as
\bea
\int\limits_{\Lambda_1}^{\Lambda_y}\frac{d\xi}{\xi}\,\ln\left(\frac{\xi + \Lambda_y}{\Lambda_y-\xi }\right) &=&
\int\limits_{\Lambda_1/\Lambda_y}^1\frac{d\xi}{\xi}\,\ln\left(\frac{1+\xi}{1-\xi }\right)
=\frac{\pi^2}{4}+{\rm Li}_2\left(-\frac{\Lambda_1}{\Lambda_y}\right)-{\rm Li}_2
\left(\frac{\Lambda_1}{\Lambda_y}\right)
\;.\label{noBCS2}
\eea
Since $0<\Lambda_1/\Lambda_y \leq 1$ this term is of order one and no BCS logarithm has 
been generated in this term.
The second term on the r.h.s.\ of Eq.\ (\ref{noBCS1}) is
\bea
 \int\limits_{\Lambda_y}^{\Lambda_0}\frac{d\xi}{\xi}\,\ln\left(\frac{\xi + \Lambda_y}
 {\xi - \Lambda_y}\right) =
\int\limits_1^{\Lambda_0/\Lambda_y}\frac{d\xi}{\xi}\,\ln\left(\frac{1+\xi}{\xi-1}\right)=
-\int\limits_1^{\Lambda_y/\Lambda_0}\frac{d\chi}{\chi}\,\ln\left(\frac{1+1/\chi}{1/\chi-1}\right)
=\frac{\pi^2}{4}+{\rm Li}_2\left(-\frac{\Lambda_y}{\Lambda_0}\right)-{\rm Li}_2
\left(\frac{\Lambda_y}{\Lambda_0}\right)
\;.
\eea
In the second step one substituted $\chi \equiv 1/\xi$ with $d\chi/\chi = -d\xi/\xi$.
Similarly to Eq.\ (\ref{noBCS2}), $0<\Lambda_y/\Lambda_0 \leq 1$. Hence, also
 this term is of order one and no BCS logarithm has been generated here, either.



\begin{thebibliography}{99}


\bibitem{bailinlove}
B.C.\ Barrois,
Nucl.\ Phys.\ B {\bf 129}, 390 (1977);
S.C.\ Frautschi,
report CALT-68-701,
{\it Presented at Workshop on Hadronic Matter at Extreme Energy
Density, Erice, Italy, Oct.\ 13-21, 1978};
for a review, see D.\ Bailin and A.\ Love,
Phys.\ Rept.\  {\bf 107}, 325 (1984).

\bibitem{RWreview}
K.\ Rajagopal and F.\ Wilczek,
arXiv:hep-ph/0011333.	

\bibitem{alfordreview}
M.G.\ Alford,
Ann.\ Rev.\ Nucl.\ Part.\ Sci.\  {\bf 51}, 131 (2001).

\bibitem{schaferreview}
T.\ Sch\"afer,
arXiv:hep-ph/0304281.

\bibitem{DHRreview}
D.H.\ Rischke,
Prog.\ Part.\ Nucl.\ Phys.\  {\bf 52}, 197 (2004).

\bibitem{Renreview}
  H.\ C.\ Ren,
  arXiv:hep-ph/0404074.

\bibitem{shovyreview}
  I.\ A.\ Shovkovy,
  Found.\ Phys.\  {\bf 35}, 1309 (2005).

\bibitem{Huang}
  M.\ Huang,
  Int.\ J.\ Mod.\ Phys.\ E {\bf 14}, 675 (2005).

\bibitem{ColPer}
J.C.~Collins and M.J.~Perry,
Phys. Rev. Lett. {\bf 34}, 1353 (1975).

\bibitem{son}
D.T.\ Son,
Phys.\ Rev.\ D {\bf 59}, 094019 (1999).

\bibitem{schaferwilczek}
T.\ Sch\"afer and F.\ Wilczek,
Phys.\ Rev.\ D {\bf 60}, 114033 (1999).

\bibitem{rdpdhr}
R.D.\ Pisarski and D.H.\ Rischke,
Phys.\ Rev.\ D {\bf 61}, 051501(R) (2000).

\bibitem{Brown1}
  W.\ E.\ Brown, J.\ T.\ Liu and H.\ C.\ Ren,
  Phys.\ Rev.\ D {\bf 61}, 114012 (2000).

\bibitem{shovkovy}
D.K.\ Hong, V.A.\ Miransky, I.A.\ Shovkovy, and L.C.R.\ Wijewardhana,
Phys.\ Rev.\ D {\bf 61}, 056001 (2000)
[Erratum-ibid.\ D {\bf 62}, 059903 (2000)].

\bibitem{Hsu}
S.D.H.\ Hsu and M.\ Schwetz,
Nucl.\ Phys.\ B {\bf 572}, 211 (2000).

\bibitem{reuter2}
  P.\ T. Reuter, PhD.\ thesis,
  arXiv:nucl-th/0602043.

\bibitem{ren}
  B.~Feng, D.~f.~Hou, J.~r.~Li and H.~c.~Ren,
  arXiv:nucl-th/0606015.

\bibitem{eliashberg}
G.\ M.\ Eliashberg, Soviet Phys.\ JETP {\bf 11}, 696 (1960).

\bibitem{schrieffer}
J.\ R.\ Schrieffer, {\it Theory of Superconductivity} (New York, W.\ A.\ Benjamin, 1964), 
D.\ J.\ Scalapino, in:  {\it Superconductivity}, ed.\ R.\ D.\ Parks, (New York, M.\ Dekker, 1969), 
p.\ 449ff.

\bibitem{vidberg}
H.\ J.\ Vidberg and J.\ Serene, J.\ Low Temp.\ Phys.\ {\bf 29}, 179 (1977).

\bibitem{mahan}
G.\ D.\ Mahan, {\it Many-Particle Physics} (Kluwer Academic/Plenum Publishers, New York, 2000). 

\bibitem{meanfield}
  R.~D.~Pisarski and D.~H.~Rischke,
  Phys.\ Rev.\ D {\bf 60}, 094013 (1999).

\bibitem{specgluon}
  R.~D.~Pisarski and D.~H.~Rischke,
  Phys.\ Rev.\ D {\bf 61}, 074017 (2000).

\bibitem{reuter}
  P.\ T.\ Reuter, Q.\ Wang and D.\ H.\ Rischke,
  Phys.\ Rev.\ D {\bf 70}, 114029 (2004)
  [Erratum-ibid.\ D {\bf 71}, 099901 (2005)].

\bibitem{Itakura}
  H.~Abuki, T.~Hatsuda and K.~Itakura,
  Phys.\ Rev.\ D {\bf 65}, 074014 (2002);
 H.\ Abuki, T.\ Hatsuda, K.\ Itakura, [arXiv:hep-ph/0206043];
  K.\ Itakura,
  Nucl.\ Phys.\ A {\bf 715}, 859 (2003).



\bibitem{lebellac}
  M.~Le Bellac and C.~Manuel,
  Phys.\ Rev.\ D {\bf 55}, 3215 (1997).

\bibitem{vanderheyden}
  B.~Vanderheyden and J.~Y.~Ollitrault,
  Phys.\ Rev.\ D {\bf 56}, 5108 (1997).
  
\bibitem{manuel}
C.\ Manuel,
Phys.\ Rev.\ D {\bf 62}, 076009 (2000).
\bibitem{manuel2}
C.\ Manuel,
Phys.\ Rev.\ D {\bf 62}, 114008 (2000).

\bibitem{rockefeller}
W.\ E.\ Brown, J.\ T.\ Liu, and H.\ C.\ Ren,
Phys.\ Rev.\ D {\bf 61}, 114012 (2000);
{\em ibid.} {\bf 62}, 054013, 054016 (2000).


\bibitem{schaferschwenzer}
  T.\ Schafer and K.\ Schwenzer,
  arXiv:hep-ph/0512309.

\bibitem{CJT}
W.\ Kohn and J.\ M.\ Luttinger
  Phys.\ Rev.\  {\bf 118}, 41-45 (1960);
 J.\ M.\ Luttinger and J.\ C.\ Ward,
  Phys.\ Rev.\  {\bf 118}, 1417 (1960);
  G.\ Baym,
  Phys.\ Rev.\  {\bf 127}, 1391 (1962);
J.M.\ Cornwall, R.\ Jackiw, and E.\ Tomboulis,
Phys.\ Rev.\ D {\bf 10}, 2428 (1974).


\bibitem{dirkselfenergy}
D.H.\ Rischke,
Phys.\ Rev.\ D {\bf 64}, 094003 (2001).

\bibitem{qwdhr}
Q.\ Wang and D.H.\ Rischke,
Phys.\ Rev.\ D {\bf 65}, 054005 (2002).


\bibitem{RDPphysicaA}
R.D.\ Pisarski,
Physica A {\bf 158}, 146 (1989).

\bibitem{LeBellac}
M.\ Le Bellac, {\it Thermal Field Theory} (Cambridge University Press, 
Cambridge, 2000).
 

\bibitem{Nielson}
N.\ Nielsen, ``Der Eulersche Dilogarithmus und seine Verallgemeinerungen,'' 
Nova Acta Leopoldina, Abh.\ der Kaiserlich Leopoldinisch-Carolinischen 
Deutschen Akad.\ der Naturforsch.\ 90, 121-212, (1909).

\end{thebibliography}
\end{document}